\documentclass[11pt,a4paper]{article}
\pdfoutput=1

\usepackage{jcappub, natbib}

\allowdisplaybreaks

\usepackage{ifthen}
\usepackage{graphicx}% Include figure files
\usepackage{dcolumn}% Align table columns on decimal point
\usepackage{bm}% bold math
\usepackage{color}
\usepackage{amsmath}
\usepackage{amssymb}
\usepackage{bbold}
\usepackage{subcaption}
\usepackage{longtable}

\newcommand{\mpc}{\, {\rm Mpc}}
\newcommand{\gpc}{\, {\rm Gpc}}

\newcommand{\hmpc}{\, h^{-1} \mpc}
\newcommand{\hgpc}{\, h^{-1} \gpc}
\newcommand{\ihmpc}{\, h\, {\rm Mpc}^{-1}}
\newcommand{\impc}{\, {\rm Mpc}^{-1}}

\newcommand{\kms}{\, {\rm km\, s}^{-1}}

\newcommand{\lya}{Ly$\alpha$\ }
\newcommand{\lyaf}{Ly$\alpha$ forest}
\newcommand{\lr}{\lambda_{\rm rest}}

\newcommand{\vk}{\mathbf{k}}
\newcommand{\vR}{\mathbf{r}}

\newcommand{\apar}{\alpha_\parallel}
\newcommand{\aperp}{\alpha_\perp}

\begin{document}

\title{Quasar-Lyman $\alpha$ Forest Cross-Correlation from BOSS DR11 : 
 Baryon Acoustic Oscillations}
\author[a,b]{Andreu Font-Ribera,}
\author[c]{David Kirkby,}
\author[d]{Nicolas Busca,}
\author[e,f]{Jordi Miralda-Escud\'{e},}
\author[b,g]{Nicholas P. Ross,}
\author[h]{An\v{z}e Slosar,}
\author[i]{James Rich,}
\author[d]{\'{E}ric Aubourg,}
\author[b]{Stephen Bailey,}
\author[j,b]{Vaishali Bhardwaj,}
\author[d]{Julian Bautista,}
\author[b]{Florian Beutler,}
\author[i]{Dmitry Bizyaev,}
\author[c]{Michael Blomqvist,}
\author[k]{Howard Brewington,}
\author[k]{Jon Brinkmann,}
\author[l]{Joel R. Brownstein,}
\author[b]{Bill Carithers,}
\author[l]{Kyle S. Dawson,}
\author[i]{Timoth\'{e}e Delubac,}
\author[k]{Garrett Ebelke,}
\author[m]{Daniel J. Eisenstein,}
\author[n]{Jian Ge,}
\author[j]{Karen Kinemuchi,}
\author[o]{Khee-Gan Lee,}
\author[k]{Viktor Malanushenko,}
\author[k]{Elena Malanushenko,}
\author[j]{Moses Marchante,}
\author[c]{Daniel Margala,}
\author[p]{Demitri Muna,}
\author[q]{Adam D. Myers,}
\author[r]{Pasquier Noterdaeme,}
\author[j]{Daniel Oravetz,}
\author[i]{Nathalie Palanque-Delabrouille,}
\author[s]{Isabelle P\^aris,}
\author[r]{Patrick Petitjean,}
\author[t]{Matthew M. Pieri,}
\author[i]{Graziano Rossi,}
\author[u,v]{Donald P. Schneider,}
\author[j]{Audrey Simmons,}
\author[w,x]{Matteo Viel,}
\author[i]{Christophe Yeche,}
\author[y]{Donald G. York}

\affiliation[a]{Institute of Theoretical Physics, University of Zurich,
        Winterthurerstrasse 190, 8057 Zurich, Switzerland}
\affiliation[b]{Lawrence Berkeley National Laboratory,
	1 Cyclotron Road, Berkeley, CA, USA}
\affiliation[c]{Department of Physics and Astronomy, University of California, 
	Irvine, CA 92697, USA}
\affiliation[d]{APC, Universit\'{e} Paris Diderot-Paris 7, CNRS/IN2P3, CEA, 
	Observatoire de Paris, 10, rue A. Domon \& L. Duquet,  Paris, France}
\affiliation[e]{Institut de Ci\`{e}ncies del Cosmos (IEEC/UB),
	Mart\'{i} i Franqu\`{e}s 1, 08028 Barcelona, Catalonia}
\affiliation[f]{Instituci\'o Catalana de Recerca i Estudis Avan\c cats,
	Passeig Lluís Companys 23, 08010 Barcelona, Catalonia}
\affiliation[g]{Department of Physics, Drexel University, 
        3141 Chestnut Street, Philadelphia, PA 19104, USA}
\affiliation[h]{Brookhaven National Laboratory,
  	Blgd 510, Upton NY 11375, USA}
\affiliation[i]{CEA, Centre de Saclay, IRFU, 91191 Gif-sur-Yvette, France}
\affiliation[j]{Department of Astronomy, University of Washington, 
	Box 351580, Seattle, WA 98195, USA}
\affiliation[k]{Apache Point Observatory and New Mexico State University, 
	P.O. Box 59, Sunspot, NM, 88349-0059, USA}
\affiliation[l]{Department of Physics and Astronomy, University of Utah, 
	115 S 1400 E, Salt Lake City, UT 84112, USA}
\affiliation[m]{Harvard-Smithsonian Center for Astrophysics,
	60 Garden Street, Cambridge, MA 02138, USA}
\affiliation[n]{Astronomy Department, University of Florida,
	211 Bryant Space Science Center, Gainesville, FL 32611-2055, USA}
\affiliation[o]{Max-Planck-Institut f\"ur Astronomie, K\"onigstuhl 17, D-69117
        Heidelberg, Germany}
\affiliation[p]{Department of Astronomy, Ohio State University, 
	Columbus, OH, 43210, USA}
\affiliation[q]{Department of Physics and Astronomy 3905, 
	University of Wyoming, 1000 East University, Laramie, WY 82071, USA}
\affiliation[r]{Institut d'Astrophysique de Paris, CNRS-UPMC, 
	UMR7095, 98bis bd Arago, 75014 Paris, France}
\affiliation[s]{Departamento de Astronom\'ia, Universidad de Chile, 
	Casilla 36-D, Santiago, Chile}
\affiliation[t]{Institute of Cosmology and Gravitation, 
	Dennis Sciama Building, University of Portsmouth, Portsmouth, 
	PO1 3FX, UK}
\affiliation[u]{Department of Astronomy and Astrophysics, 
	The Pennsylvania State University, University Park, PA 16802}
\affiliation[v]{Institute for Gravitation and the Cosmos, 
	The Pennsylvania State University, University Park, PA 16802}
\affiliation[w]{INAF, Osservatorio Astronomico di Trieste, 
	Via G. B. Tiepolo 11, 34131 Trieste, Italy}
\affiliation[x]{INFN/National Institute for Nuclear Physics, 
	Via Valerio 2, I-34127 Trieste, Italy}
\affiliation[y]{Deptartment of Astronomy and Astrophysics and The Fermi 
	Institute, The University of Chicago, 5640 South Ellis Avenue, 
	Chicago, IL 60615, USA}

\emailAdd{afont@lbl.gov}

\abstract{
We measure the large-scale cross-correlation of quasars with the \lya forest 
absorption, using over 164,000 quasars from Data Release 11 of the SDSS-III 
Baryon Oscillation Spectroscopic Survey.
We extend the previous study of roughly 60,000 quasars from Data Release 9 
to larger separations, allowing a measurement of the Baryonic Acoustic 
Oscillation (BAO) scale along the line of sight 
$c/(H(z=2.36) ~ r_s) = 9.0 \pm 0.3$
and across the line of sight 
$D_A(z=2.36)~ / ~ r_s = 10.8 \pm 0.4$, 
consistent with CMB and other BAO data. 
Using the best fit value of the sound horizon from Planck data 
($r_s=147.49 \mpc$), we can translate these results to a measurement of the 
Hubble parameter of $H(z=2.36) = 226 \pm 8 \kms \impc$ and of the angular 
diameter distance of $D_A(z=2.36) = 1590 \pm 60 \mpc$.
The measured cross-correlation function and an update of the code to fit the 
BAO scale (baofit) are made publicly available.
}


\maketitle

\section{Introduction}

Fifteen years ago, two independent studies of the luminosity distance of 
type Ia supernovae (\cite{RIFIET98},\cite{PEADET99}) showed that the Universe 
was undergoing an accelerated expansion. In order to explain such
an unintuitive result, different authors have suggested the need for a 
cosmological constant in Einstein's equations of general relativity, 
more profound modifications of the gravitational theory, or the presence of a 
new energy component usually referred to as \textit{dark energy}. 

Following this discovery, different cosmological probes have provided a wealth
of new data, allowing us to constrain the cosmological parameters of the model 
at a few-percent level. The simplest possible solution, a flat universe with a 
cosmological constant, is able to explain all current data 
\cite{2013arXiv1303.5076P}, and ongoing and future cosmological surveys will 
continue to reduce the errorbars of these measurements and place even 
more stringent constraints on the models.

There are different observational probes that can measure the history of the
accelerated expansion (see \cite{2012arXiv1201.2434W} for a review). 
One is the measurement of the Baryon Acoustic Oscillation (BAO) 
scale on the clustering of any tracer of the density field, which can be used 
as a cosmic ruler to study the geometry of the Universe 
\cite{2003ApJ...598..720S}. This technique has gained considerable attention 
during the last decade, and the list of BAO measurements is rapidly increasing.

In theory, any tracer of the large-scale matter distribution can be used to 
measure BAO. Even though the first measurements came from the clustering of 
low redshift galaxies (Sloan Digital Sky Survey \cite{2005ApJ...633..560E} 
at $0.2<z<0.4$, Two-degree-Field Galaxy Redshift Survey 
\cite{2005MNRAS.362..505C} at $0.1<z<0.2$), and the tightest constraints are
obtained from intermediate redshift galaxy surveys 
(WiggleZ Dark Enery Survey \cite{2011MNRAS.418.1707B} at $0.4<z<0.8$, 
Baryon Oscillation Spectroscopic Survey \cite{2012MNRAS.427.3435A} at 
$0.4<z<0.7$), there are also a variety 
of undergoing or planned surveys that aim to measure BAO at higher redshift 
from the clustering of x-ray sources (eROSITA \cite{2012arXiv1209.3114M}), 
21cm emission (Canadian Hydrogen Intensity Mapping Experiment 
\footnote{http://chime.phas.ubc.ca/}, Baryon Acoustic Oscillation Broadband 
and Broad-beam Array \cite{2013AJ....145...65P}), \lya emitting galaxies 
(Hobby-Eberly Telescope Dark Energy Experiment \cite{2008ASPC..399..115H},
Dark Energy Spectroscopic Instrument \cite{2009arXiv0904.0468S}) and quasars 
(Extended Baryon Oscillation Spectroscopic Survey \cite{eBOSS}). 
 
The spatial distribution of neutral hydrogen, as traced by the Lyman-$\alpha$
forest (\lya forest) can also be used to measure BAO. The first 
measurement of the three-dimensional large-scale structure of \lya absorption 
was presented in \cite{2011JCAP...09..001S}, using over 14,000 spectra from 
the first year of BOSS. This study was extended using approximately 
50,000 quasar spectra from the ninth data release of SDSS 
(DR9, \cite{2012ApJS..203...21A}), and the first detections of the BAO 
at $z=2.4$ were presented in \cite{2013A&A...552A..96B}, 
\cite{2013JCAP...04..026S} and \cite{2013JCAP...03..024K}. 

Using the same set of spectra, \cite{2013JCAP...05..018F} presented an 
analysis of the large scale cross-correlation of quasars and the \lya
absorption. In this analysis the cross-correlation was clearly detected
up to separations of $r \sim 70 \hmpc$, and was accurately described by 
a linear bias and redshift-space distortion theory for comoving separations 
$r > 15 \hmpc$. Their measurement of the 
quasar bias of $b_q = 3.64 \pm 0.14$ was fully consistent with measurements 
of the quasar auto-correlation function at the same redshift, e.g.,
\cite{2012MNRAS.424..933W}.

In this paper, we use over 164,000 quasars from the eleventh data release of 
SDSS (DR11, which will be publicly released at the end of 2014 together with 
DR12) to extend the measurement of the cross-correlation to larger 
separations, and we present an accurate determination of the BAO scale in 
cross-correlation at high-redshift ($z\sim2.4$).

Throughout, we use the fiducial cosmology 
($\Omega_m = 0.27$, $\omega_b = 0.0227$, $h=0.7$, $\sigma_8=0.8$, $n_s=0.97$) 
that was used in the previous \lya BAO measurements 
(\cite{2013A&A...552A..96B}, \cite{2013JCAP...04..026S},
\cite{2013JCAP...03..024K}). We use the publicly available code CAMB 
\cite{2000ApJ...538..473L} to compute the comoving distance to the sound 
horizon at the redshift at which baryon-drag optical depth equals unity, 
$z_{drag}=1059.97$, and obtain a value of $r_s = 149.72 \mpc$. 
Some previous BAO studies have used the equations from \cite{EIHU98} to 
compute the value of $r_s$, resulting in a few percent difference with 
respect to the value computed with CAMB, as discussed in 
\cite{2013arXiv1303.5076P}. 

We start by introducing our data sample in section \ref{sec:data}. 
In section \ref{sec:cross} we present our measurement of the quasar-\lya 
cross-correlation and summarize our analysis method. 
In section \ref{sec:bao} we describe our fits of the BAO scale, present our 
main results, and test possible systematic effects. 
In section \ref{sec:conc} we compare our results with previous BAO 
measurements at similar redshifts. 

\section{Data Sample}
\label{sec:data}

In this section we describe the data set used in this study, and present 
a series of references for further details. 

The eleventh Data Release (DR11) of the SDSS-III Collaboration (
\cite{2011AJ....142...72E}, % Eisenstein++ 2011, SDSS-III
\cite{2012AJ....144..144B}, % Bolton++ 2012, pipeline
\cite{1998AJ....116.3040G}, % Gunn++ 1998, SDSS camera
\cite{2006AJ....131.2332G}, % Gunn++ 2006, SDSS telescope
\cite{2013AJ....146...32S}, % Smee++ 2012, SDSS/BOSS spectrographs
\cite{2000AJ....120.1579Y} % York++ 2000, SDSS-I/II technical summary
) contains all spectra obtained during the first four years of the 
Baryon Oscillation Spectroscopic Survey (BOSS, \cite{2013AJ....145...10D}), 
including spectra of 238,978 visually confirmed quasars. 
The quasar target selection used in BOSS is summarized in 
\cite{2012ApJS..199....3R}, and combines different targeting methods described
in \cite{2010A&A...523A..14Y}, \cite{2011ApJ...743..125K}, and
\cite{2011ApJ...729..141B}.

In this study we measure the cross-correlation of two tracers of the 
underlying density field: the number density of quasars and the \lya absorption
along a set of lines of sight. We will use the term ``quasar sample'' to 
refer to the quasars used as tracers of the density field, and the term 
``\lya sample'' to refer to those quasar lines of sight where the \lya 
absorption is measured.

\subsection{Quasar sample}

We use a preliminary version of the DR11Q quasar catalog, an updated version 
of the DR9Q catalog presented in \cite{2012A&A...548A..66P}. This catalog 
contains a total of 238,978 visually confirmed quasars, distributed in an area 
of 8976 square degrees in two disconnected parts of the sky: the 
South Galactic Cap (SGC) and the North Galactic Cap (NGC). 
To avoid repeated observations of the same object, we only use quasars in the
catalog that have the SPECPRIMARY flag \cite{2013AJ....145...10D}.

The performance of the BOSS spectrograph rapidly deteriorates at wavelengths 
bluer than $\lambda \sim 3650 {\, \rm\AA }$, corresponding to a \lya absorption
of $z=2$; this wavelength sets the lower limit in the redshift range used in 
this study. Since the number of identified quasars drops rapidly at high 
redshift, we restrict our quasar sample to those with redshifts in the 
range $2.0 \le z_q \le 3.5$. This redshift constraint reduces the number of 
quasars used as tracers of the density field to 164,017. In this paper we 
use the Z\_VI redshift estimate from \cite{2012A&A...548A..66P}.

\subsection{\lya sample}

Not all spectra from the quasar sample are included in our \lya sample. 
We first drop the spectra from quasars with a redshift lower 
than $z_q = 2.15$, since only a small part of their \lya forest can be 
observed with the BOSS spectrograph. This choice reduces the number of 
spectra to 153,496. 

During the visual inspection, Broad Absorption Line quasars (BAL) are 
identified \cite{2012A&A...548A..66P}. We discard any spectra from BAL 
quasars, reducing the number of spectra to 136,431. 
We finally exclude spectra with less than 150 pixels covering the \lya forest, 
further reducing the number of spectra used in the \lya sample to 130,825. 
We use the same definition of the \lya forest as in \cite{DR11BAO}, which 
contains all pixels in the rest-frame wavelength range
$1040 {\, \rm\AA } \le \lr \le 1200 {\, \rm\AA }$. 

In the left panel of figure \ref{fig:sample} we show the redshift distribution 
of objects in the quasar sample (red in the NGC, blue in the SGC) and the 
distribution of these quasars that have spectra included in the \lya sample 
(green in the NGC, purple in the SGC). 

We use an updated (DR11) version of the Damped Lyman $\alpha$ system (DLA) 
catalogue from \cite{2012A&A...547L...1N}, based on DLA profile recognition as 
described in \cite{2009A&A...505.1087N}, to mask the central part of DLA 
in the spectra (up to a transmitted flux fraction of $F<0.8$), and correct 
the rest of the spectra using the inferred Voigt profile. 

We correct the noise estimate from the pipeline using the method described 
in \cite{2013A&A...552A..96B}. 
Following the same reference, we rebin our spectra by averaging the flux over 
three adjacent pipeline pixels. These new pixels have a width of 
$210 \kms$, $\sim 2 ~\hmpc$ at the redshift of interest, much smaller than 
the minimum separation in which we are interested ($r > 40 \hmpc$), and 
smaller than the width of the BAO peak ($\Delta r \sim 25 \hmpc$). 
We will use the term ``pixel'' to refer to these rebinned pixels.

\subsection{Independent sub-samples}

\begin{figure}
 \begin{center}
  \includegraphics[scale=0.3]{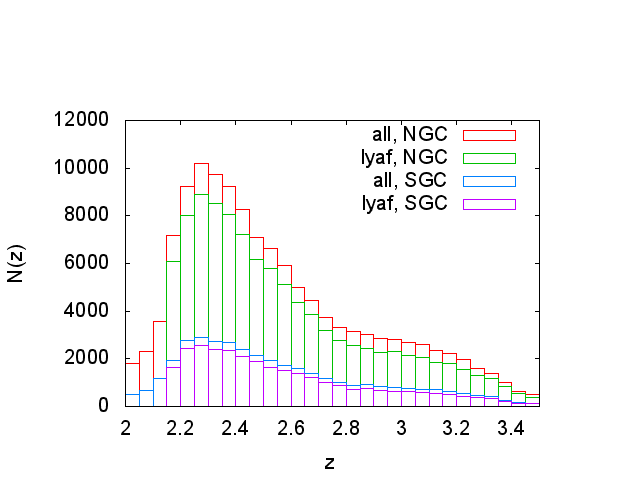}
  \includegraphics[scale=0.3]{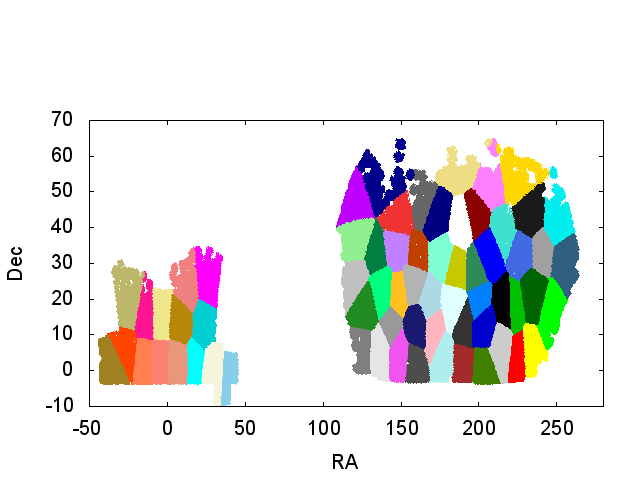}
 \end{center}
 \caption{Left panel: Redshift distribution of the 164,017 quasars used as 
  density tracers (red in NGC, blue in SGC), 
  and of the 130,825 quasars with \lya spectra (green in NGC, purple in SGC). 
  Right panel: DR11 footprint in J2000 equatorial coordinates with the 66 
  sub-samples indicated in different colors.}
 \label{fig:sample}
\end{figure}

In section \ref{sec:cross} we explain the method to estimate the covariance
matrix of our measurement. To reduce the required computing time, the 
survey is split into 66 sub-samples with a similar 
number of quasars and combine the measurement of each sub-sample assuming 
that they are independent. The right panel of figure \ref{fig:sample} shows
the different sub-samples, 51 of them in the north galactic cap (NGC) and 
15 in the south (SGC). 

These sub-samples are also used in section \ref{sec:bao} to compute 
bootstrap errors on the best fit parameters. 
Since we are interested in scales smaller than $150 \hmpc$ and the 
typical area of these sub-samples is roughly $(800 \hmpc)^2$ 
($140 \, {\rm deg}^2$ at $z=2.4$), the assumption that the sub-samples are 
independent is justified. 

\section{Cross-correlation}
\label{sec:cross}

In this section we briefly describe the method used to measure the 
cross-correlation and its covariance matrix, referring the reader to previous 
publications for a detailed explanation (\cite{2012JCAP...11..059F},
\cite{2013JCAP...05..018F}), and present the measured cross-correlation.

\subsection{Continuum fitting}

The first step necessary to estimate the Lya transmitted flux fraction
$F(\lambda)=e^{-\tau(\lambda)}$ from a set of pixels with flux $f(\lambda)$ 
is estimating the quasar continuum, $C_q(\lambda)$,
\begin{equation}
 F(\lambda) = \frac{f(\lambda)}{C_q(\lambda)} ~.
\end{equation}

Among various approaches for the determination of the continuum available
in the literature, we use \textit{Method 2} in \cite{2013A&A...552A..96B}, 
which is also being used for the analysis of the BAO Lya autocorrelation 
for DR11 \cite{DR11BAO}. This method assumes that all quasars have the same 
continuum $\bar C(\lr)$, except for a linear multiplicative function that 
varies for each quasar:
\begin{equation}
C_q(\lambda)=(a_q + b_q\lambda ) \bar C (\lr) ~,
\end{equation}
where $a_q$ and $b_q$ are fitted to match an assumed probability distribution
function (PDF), as explained in \cite{2013A&A...552A..96B}.

The construction of the continuum is a critical step for those \lya studies 
focused on the line of sight power spectrum or in the flux PDF, where errors 
in the continuum fitting can systematically bias the results. 
In three dimensional clustering measurements of \lya absorption, one would 
expect that the continuum fitting errors in different lines of sight are
uncorrelated, getting rid of any potential bias in the measurement. 
However, as noted by \cite{2011JCAP...09..001S}, if the continuum of each
quasar is rescaled in order to match an external mean transmission or flux PDF,
the errors in the continuum fitting will be correlated with large scale 
density fluctuations. We discuss this issue further in section \ref{ss:MTC}. 

In section \ref{sec:bao} (table \ref{tab:bao}) we present an alternative 
analysis using a different method to fit the continua, based on the 
Principal Component Analysis (PCA) presented in \cite{2012AJ....143...51L}, 
and show that the measurement of the BAO scale is in very good agreement 
with our fiducial analysis.

\subsection{From flux to $\delta_F$}

Using the continua described above, we now measure the mean transmitted flux 
fraction $\bar F (z)$, also known as the ``mean transmission''. The redshift of 
absorption $z$ is related to the observed wavelength $\lambda$ by the 
\lya transition line $\lambda_\alpha=1216 {\, \rm\AA } = \lambda / (1+z)$ .
We measure the mean transmission as a function of redshift, in $N_z=300$ 
bins between $1.9 \le z \le 3.4$,
\begin{equation}
 \bar F (z) = \left< \frac{f(\lambda)}{C_q (\lambda)}  \right> ~, 
\end{equation}
where the average is computed over all pixels in each redshift bin.
The \lya absorption fluctuation is then defined as 
\begin{equation}
 \delta_F = \frac{f}{C_q \bar F} - 1 ~.
 \label{eq:delta_f}
\end{equation}

As noted in  \cite{2013A&A...552A..96B}, there are sharp features in the 
measurement of the mean transmission due to imperfections in the calibration 
vector of the BOSS data reduction pipeline.
We do not expect this error to bias our results on the quasar-\lya
cross-correlation because it should be corrected when the quasar spectra
are divided by the measured mean transmission, and any residual errors are 
not expected to correlate with the quasar detection efficiency that 
varies across the BOSS survey area.

\subsection{Estimator and covariance matrix}

We estimate the cross-correlation $\xi_A$ between quasars and \lya absorption, 
in a bin $r_A$, employing the same method that was used in previous analyses 
of cross-correlations in BOSS (\cite{2012JCAP...11..059F}, 
\cite{2013JCAP...05..018F}):
\begin{equation}
 \hat\xi_{A} = \frac{\sum_{i \in A}  w_i\, \delta_{Fi}}{\sum_{i \in A} w_i} ~,
 \label{eq:xiA}
\end{equation}
where the sum is over all pixels $i$ that are at a separation $r_i$ in bin $A$ 
from a quasar, and where the weights $w_i$ are 
computed independently at each pixel from the pipeline noise variance and 
assuming a model for the intrinsic \lya absorption variance
(equation 3.10 in \cite{2012JCAP...11..059F}).

The covariance matrix of the correlation measurements in two bins $A$ and $B$,
$\tilde C_{AB}$, is too large to be computed using resampling techniques 
or from synthetic data sets.
Instead we use an analytical estimate similar to the method used in \lya
auto-correlation analyses (\cite{2011JCAP...09..001S},\cite{DR11BAO}), which
was first applied to cross-correlations studies in \cite{2012JCAP...11..059F} :
\begin{equation}
 \tilde C_{AB} \equiv \left< \hat\xi_{A} \hat\xi_{B} \right> -
 \left< \hat\xi_{A} \right> \left< \hat\xi_{B} \right> 
   = \frac{\sum_{i \in A} \sum_{j \in B}  w_i ~ w_j ~ C^F_{ij}}
          {\sum_{i \in A} w_i ~ \sum_{j \in B}  w_j} ~,
 \label{eq:covar}
\end{equation}
where $C^F_{ij} = \left< \delta_{Fi}\delta_{Fj}\right>$ is the correlation
of the Lyman alpha fluctuations $\delta_F$ measured in pixels $i$ and $j$, 
separated in redshift space by $\mathbf{r}_{ij}$, and includes both the
cosmological signal and the contribution from instrumental noise when $i=j$. 
In this study, we further simplify the calculation by ignoring the 
correlation among \lya pixels in different lines of sight. 

This method assumes Gaussian errors and ignores the contribution from
cosmic variance, a reasonable approximation given the large volume of the
survey and its sparse sampling. We also assume that the variance in the
quasar density field is dominated by shot noise, which is justified on the
same grounds. It is useful to examine these assumptions in Fourier space
(see appendix \ref{app:forecast} for an extended discussion on this). 
The variance in the measurement of the cross-power spectrum $P_{qF}(\vk)$ 
for a single Fourier mode $\vk$ can be approximated by :
\begin{equation}
  var \left( P_{qF}(\vk) \right) = 
      P_{qF}(\vk)^2 + \left(P_{qq}(\vk) + n_q^{-1} \right) 
      \left(P_{FF}(\vk) + P^{1D}(k \mu) ~ n_{eff}^{-1} \right) ~,
\end{equation}
where $n_q$ is the quasar density, $P_{qq}$ the quasar auto-power spectrum, 
$P_{FF}$ the \lya auto-power spectrum, $P^{1D}$ the line of sight \lya 
power spectrum and $n_{eff}$ the effective density of \lya lines of sight
as defined in \cite{2011MNRAS.415.2257M}. Ignoring cosmic variance is 
equivalent to removing the term $P_{qF}^2$, while assuming that quasars are
shot-noise dominated is equivalent to removing the $P_{qq}$ term. These 
approximations are supported by the analysis of the various terms in this 
equation presented in \ref{app:forecast}.

We use 16 bins of constant width $10 \hmpc$ in transverse separation 
$r_\perp$, up to a maximum separation of $r_\perp < 160 \hmpc$. 
Since $r_\parallel$ can be positive (pixel behind the quasar) 
or negative (pixel in front of the quasar) we use 32 bins in 
$r_\parallel$ with the limits $-160 \hmpc < r_\parallel < 160 \hmpc$. 
We use a single bin in redshift, ranging from $2.0 < z < 3.4$.
Other BAO studies that measure the correlation in multipoles, or grids 
defined in the ($r=\sqrt{r_\parallel^2+r_\perp^2}$, $\mu=r_\parallel/r$) plane, 
use narrower bins in order to better resolve the BAO peak. 
Coarser bins can be used in studies where the correlation is measured in 
the ($r_\parallel$,$r_\perp$) plane, since each point corresponds to a 
different value of $r$. For instance, we cover 48 different values of $r$ 
in the range $90 \hmpc < r < 120 \hmpc$.

As discussed in \ref{ss:subsamples}, the cross-correlations and their 
covariance matrices are measured in 66 sub-samples (shown in figure 
\ref{fig:sample}), $\mathbf{\hat\xi_\alpha}$ and $\tilde C_{\alpha}$.
Assuming that these are independent, the optimal way to combine them is:
\begin{equation}
 \tilde C^{-1} = \sum_\alpha \tilde C^{-1}_\alpha ~, 
\qquad
 \mathbf{\hat\xi} = \tilde C \sum_\alpha \tilde C^{-1}_\alpha
 \mathbf{\hat\xi_\alpha} ~. 
 \label{eq:total_xi}
\end{equation}

When measuring the correlation in one of the sub-samples we only use \lya 
pixels from spectra in that given part of the sky. However, we cross-correlate
the absorption in these pixels not only with quasars from the sub-sample, but 
also with quasars in the neighboring sub-samples. We are therefore not losing
any interesting quasar-pixel pairs, at the expense of adding a small correlation
between the different measurements.

\subsection{Measured cross-correlation}
\label{ss:MTC}

\begin{figure}
 \begin{center}
  \includegraphics[scale=0.15]{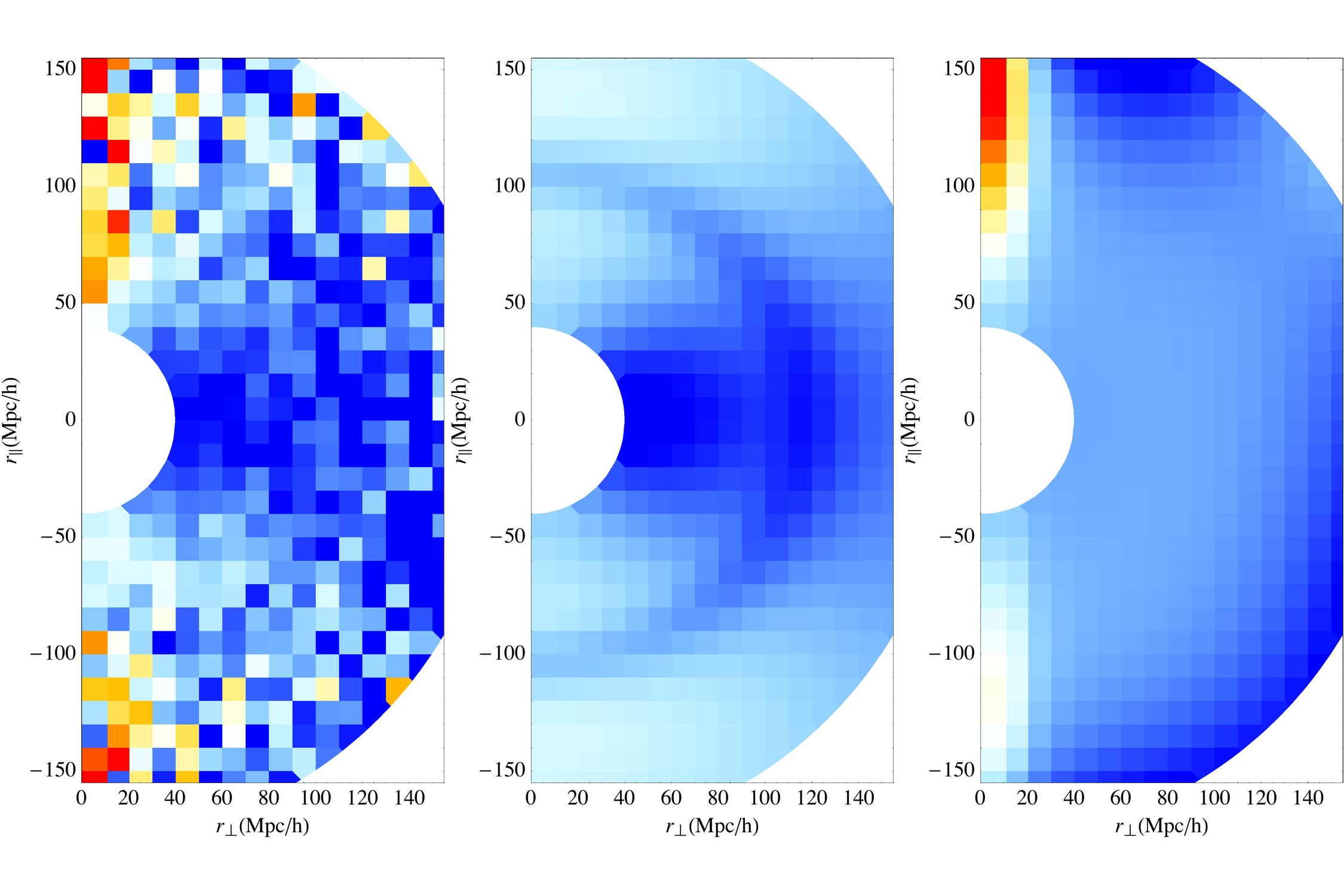}
 \end{center}
 \caption{Left panel: Observed quasar-\lya cross-correlation, 
  $r^2 \xi(r_\parallel,r_\perp)$, as a function of line of sight 
  ($r_\parallel$) and transverse ($r_\perp$) separations. Mid panel: 
  Cosmological contribution to the best fit model (see next section). 
  Right panel: contribution of the broadband distortion to the best fit model. 
  The fit to the observed cross-correlation is the sum of the middle and 
  right panels. As discussed in the text, the asymetric nature of the 
  broadband distortion can be explained by the continuum fitting method.
}
 \label{fig:contours}
\end{figure}

 The quasar-\lya cross-correlation that is obtained with the method
just described is plotted in the left panel of figure \ref{fig:contours}. 
The model that we use to fit its functional form has two components: first, 
the theoretical cross-correlation function in the absence of systematics, 
and second, a broadband function that models systematic distortions that
are introduced into the measurement. This will be described in detail in
section \ref{sec:bao}, but it is useful to discuss now the general reason to 
fit a model with these two terms.

The middle panel in figure \ref{fig:contours} shows the cosmological 
component of the best fit model, and the right panel is the broadband 
distortion part. The fit to the observed cross-correlation is the sum 
of the functions in the middle and right panels. The shape of the broadband 
distortion and its asymetric nature can be explained by our method to 
determine the quasar continuum, which involves fitting a multiplicative 
function $a_q+b_q \lambda$ to each spectrum to match an assumed PDF of the
transmission $F$. This effectively removes large-scale power in the
observed \lya forest: roughly speaking, the mean value and the
gradient of the large-scale density fluctuations over the line of sight
of each \lya spectrum are removed by the continuum fitting operation.

The distortion effect that this introduces on three-dimensional correlation
measurements was first discussed in the context of the \lya auto-correlation 
in Appendix A of \cite{2011JCAP...09..001S}. The corresponding distortion on
the cross-correlation was considered in \cite{2012JCAP...11..059F}, where it 
was modeled and computed in terms of the quasar redshift distribution and the
interval of the observed \lya forest spectra. This expected distortion,
plotted in figure 17 of \cite{2012JCAP...11..059F}, is a strong function of 
$r_\perp$ and a weaker function of $r_\parallel$, and is asymetric under a 
sign change of $r_\parallel$ because the average quasar redshift is higher 
than the average \lya pixel redshift.

An analytical prescription to correct 
for this distortion in the fitted theoretical model was presented in 
\cite{2012JCAP...11..059F}, valid for the simpler continuum fitting method 
that was used there. This was crucial for that work and in 
\cite{2013JCAP...05..018F}, where the goal was to accurately measure the
full shape of the cross-correlation to obtain the bias and redshift
distortion parameters.

In this paper, our goal is to measure the position of the BAO peak without 
any dependence on possible systematics in our modeling of the broadband shape 
of the cross-correlation. We therefore decide not to apply any correction to 
the theoretical model. Instead, a broad-band term is added to the 
cross-correlation with enough free parameters to absorb a generic smooth 
distortion, as explained in section \ref{sec:bao}. This approach, also used 
in the recent BAO measurements from the \lya auto-correlation
(\cite{2013A&A...552A..96B}, \cite{2013JCAP...04..026S}, 
\cite{2013JCAP...03..024K}), relies on 
the narrowness of the BAO peak, which decouples its position from the 
broadband shape. Unfortunately, this degrades our ability to measure the 
bias and redshift distortion parameters because they are affected by the 
broadband model.

\section{Fitting the BAO Scale}
\label{sec:bao}

In this section we describe the method used to measure the scale of the 
Baryon Acoustic Oscillations (BAO) from the measured cross-correlations, and 
present our main results. We conclude with a detailed analysis of possible
sources of systematic errors.

\subsection{BAO model} 

We adapted the publicly available fitting code baofit 
\cite{2013JCAP...03..024K} to work with cross-correlations. The code can
be downloaded from the URL in footnote 
\footnote{http://darkmatter.ps.uci.edu/baofit/},
together with the measured cross-correlation and its covariance matrix, as 
described in appendix \ref{app:access}.

A detailed description of the fitting code can be found in 
\cite{2013JCAP...03..024K}; here we only summarize the main points and 
highlight the differences between fitting the \lya auto-correlation and the
quasar-\lya cross-correlation.

We model the measured cross-correlation $\xi(r_\parallel,r_\perp)$ as a sum 
of the \textit{cosmological} correlation $\xi_{\rm cosmo}$ and a broadband 
distortion term $\xi_{\rm bb}$ due primarily to continuum fitting 
(as discussed in section \ref{ss:MTC})
\begin{equation}
  \xi (r_\parallel, r_\perp) 
	= \xi_{\rm cosmo} (r_\parallel,r_\perp,\apar,\aperp) 
		+ \xi_{\rm bb} (r_\parallel,r_\perp) ~.
\end{equation}
The quantity $\xi_{\rm cosmo}$ is described as a sum of two terms
\begin{equation}
  \xi_{\rm cosmo}(r_\parallel,r_\perp) 
	= \xi_{\rm smooth} (r_\parallel, r_\perp) +
  a_{\rm peak} \cdot \xi_{\rm peak} (r_\parallel \apar, r_\perp \aperp)~,
\end{equation}
where $a_{\rm peak}$ controls the amplitude of the BAO peak. 
The correct $\xi_{\rm cosmo}$ in the Cold Dark Matter standard model is 
obtained only for $a_{\rm peak}=1$, which we use in all of our analyses 
except when we want to test the consistency of our results with the 
prediction for the peak amplitude (row labeled AMP in table \ref{tab:bao}).

The main goal of this study is to measure the scale of BAO relative to the 
fiducial cosmological model, along the line of sight $\apar$ and 
across the line of sight $\aperp$:
\begin{equation}
 \apar = \frac { \left[D_H(\bar z)/r_s\right] }
			{\left[D_H(\bar z)/r_s\right]_{\rm fid}} ~, \qquad
 \aperp = \frac { \left[D_A(\bar z)/r_s\right] }
			{\left[D_A(\bar z)/r_s\right]_{\rm fid}} ~, 
 \label{eq:alpha}
\end{equation}
where $r_s$ is the sound horizon, $D_H(z) = c/H(z)$ is the Hubble distance,
and $D_A(z)$ the comoving angular diameter distance. The mean redshift of our
measurement is $\bar z=2.36$. 
 
The scale factors ($\apar$,$\aperp$) only appear in the \textit{peak}
part of the correlation, to ensure that no information comes from the 
broadband shape. A detailed description of the decomposition of the 
cosmological signal into a \textit{peak} and a \textit{smooth} component 
can be found in \cite{2013JCAP...03..024K}.

\subsubsection{Theoretical model for the cross-correlation}

We model the \textit{cosmological} correlation 
$\xi_{\rm cosmo}(\vR)$ 
as the 3D Fourier transform of the cross-power spectrum $P_{qF}(\vk)$:
\begin{equation}
 P_{qF}(\vk) = b_q \left[1+\beta_q \mu_k^2 \right] 
           b_F \left[1+\beta_F \mu_k^2 \right] \, P(k) ~,
 \label{eq:Px}
\end{equation}
where $b_q$ is the linear bias of quasars, $b_F$ the linear bias of \lya 
forest, and $\beta_q$ and $\beta_F$ the redshift space distortion parameters 
for quasars and \lya forest. The matter power spectrum is $P(k)$, which 
includes the non-linear broadening of the peak \cite{2013JCAP...03..024K} 
and $\mu_k$ is the cosine of the angle between the Fourier mode vector $\vk$
and the line of sight.

Following \cite{2013JCAP...05..018F} we leave two of the four bias parameters
free ($b_q$ and $\beta_F$) and derive the other two from them, using the 
well-constrained combination $b_F(1+\beta_F)=-0.336 \pm 0.03$ 
\cite{2011JCAP...09..001S} and the Kaiser relation $b_q \beta_q=f(\Omega_m)$ 
\cite{1987MNRAS.227....1K}, where $f(\Omega_m$) is the logarithmic growth 
rate of structure. Note that the same relation does not apply to the \lya 
forest (e.g., \cite{2011JCAP...09..001S}). These values of the bias 
parameters are defined at $z=2.25$, and we translate them to our mean 
redshift $\bar z=2.36$ assuming that only $b_F$ evolves with redshift, 
following $ \left( b_F(z) g(z) \right)^2 \propto (1+z)^{3.8}$, where $g(z)$ 
is the linear growth factor (as discussed in \cite{2011JCAP...09..001S}).

\subsubsection{Quasar redshift errors} 

Determining precise quasar redshifts is a difficult task. As noted in 
\cite{2013JCAP...05..018F}, quasar redshift errors have two main effects 
on the cross-correlation: 
a) the r.m.s.\ in the quasar redshift estimates ($\sigma_z \sim 500 \kms$) 
smooths the cross-correlation along the line of sight (with an equivalent 
effect on the quasar auto-correlation, \cite{2012MNRAS.424..933W});
b) a systematic offset in the BOSS redshift estimates shifts the 
cross-correlation along the line of sight by a non-negligible amount 
$\Delta_z \sim - 180 \kms$ (see \cite{2013JCAP...05..018F}).

Since we restrict our analysis to large separations ($r>40 \hmpc$), we do not
expect quasar redshift errors to have a significant impact on our fits. We 
leave $\Delta_z$ as a free parameter in all our fits, presenting our 
results after marginalizing over it. We do not include an explicit $\sigma_z$ 
parameter since it would be highly degenerate with the non-linear broadening 
model.

\subsubsection{Broadband distortion}
\label{ss:BB}

All BAO analyses to date have used a broadband model that parameterizes each
multipole as a function of $r$, or a parameterization as a function of 
($r$,$\mu$) as in \cite{2013JCAP...04..026S}. However, the shape of the 
distortion discussed in section \ref{ss:MTC} that is introduced into the 
cross-correlation by the continuum fitting operation is better separated in 
terms of the ($r_\parallel$, $r_\perp$) coordinates, as inferred from the 
analysis that was presented in \cite{2012JCAP...11..059F} 
(see their figure 17). In this figure, one can see that the distortion
decreases rapidly with $r_\perp$, and it has a non-trivial $r_\parallel$
dependence. Therefore, we use the following parameterization for the 
broadband distortion model:
\begin{equation}
 \label{eq:bb}
 \xi_{bb}(r_\parallel,r_\perp) = \sum_{i=i_{min}}^{i_{max}} 
	 \sum_{j=j_{min}}^{j_{max}} b_{ij} ~ r_\parallel^i ~ r_\perp^j ~,
\end{equation}
where the sums are understood to be over consecutive integers, and they
go from $i_{\rm min}=0$ to $i_{\rm max}=2$, and from $j_{\rm min}=-3$ to
$j_{\rm max}=1$ in our fiducial analysis. The dependence of our results
on the broadband distortion model is discussed in section \ref{ss:syst}.

\subsection{BAO fits}

Our fiducial BAO fit is performed over the separation range 
$40 \hmpc < r < 180 \hmpc$ using a broadband model with 
($i_{min}=0$, $i_{max}=2$, $j_{min}=-3$, $j_{max}=1$). 
The total number of bins included is 440, and the number of free parameters 
is 20: $\apar$, $\aperp$, $\beta_F$, $b_q$, $\Delta_z$ and the 15 
parameters $b_{ij}$ in our broadband distortion model. 
In table \ref{tab:bao} we present the best fit values for our fiducial 
analysis, and for a series of illustrative alternative analyses:
an isotropic BAO analysis (ISO) imposing $\alpha \equiv \apar = \aperp$; 
a no-wiggles fit (NW) with $a_{\rm peak}=0$; a fit allowing the amplitude 
of the peak $a_{\rm peak}$ to vary (AMP); and a fit using a different method 
to fit the continua based on a Principal Component Analysis 
(PCA, \cite{2012AJ....143...51L}).

The BAO peak position is significantly measured to an accuracy better than 
$\sim$ 4\% both along and across the line of sight directions. 
The measured amplitude of the BAO peak is consistent with the expected in 
our fiducial model. 

\begin{table}[t!]
 \centering
 \small
  \begin{tabular}{c|ccccccc}
   &  $\beta_F$  & $b_q$  & $a_{\rm peak}$ & $\alpha$  &  $\apar$  &  $\aperp$  & $\chi^2$ (d.o.f) \\
\hline
FID   & $1.09 \pm 0.29 $  & $3.02 \pm 0.22 $ & - 		& - 			& $1.042 \pm 0.034 $  	& $0.930 \pm 0.036 $  	& 426.4 (420) \\ 
\hline	
ISO   & $1.09 \pm 0.30 $  & $3.00 \pm 0.22 $ & - 		& $0.988 \pm 0.022 $  	& - 			& - 			& 429.5 (421) \\ 
NW    & $1.13 \pm 0.34 $  & $2.71 \pm 0.21 $ & - 		& - 			& - 			& - 			& 448.5 (422) \\ 
AMP   & $1.14 \pm 0.32 $  & $3.02 \pm 0.22 $ & $1.15 \pm 0.24 $ & - 			& $1.041 \pm 0.033 $  	& $0.933 \pm 0.035 $  	& 426.0 (419) \\ 
PCA  & $1.54 \pm 0.42 $   & $3.14 \pm 0.22 $ & - 		& - 			& $1.052 \pm 0.035 $  	& $0.922 \pm 0.044 $  	& 474.7 (420) \\ 
 \end{tabular}
 \caption{Best fit parameters for different analyses : 
  fiducial BAO fit (FID), istropic BAO fit (ISO),
  non-BAO fit (NW), fit with free amplitude (AMP)
  and using a continuum fitting method based on a Principal 
  Component Analysis (PCA, \cite{2012AJ....143...51L}). 
 }
 \label{tab:bao}
\end{table}

\begin{figure}
 \begin{center}
  \includegraphics[scale=0.6]{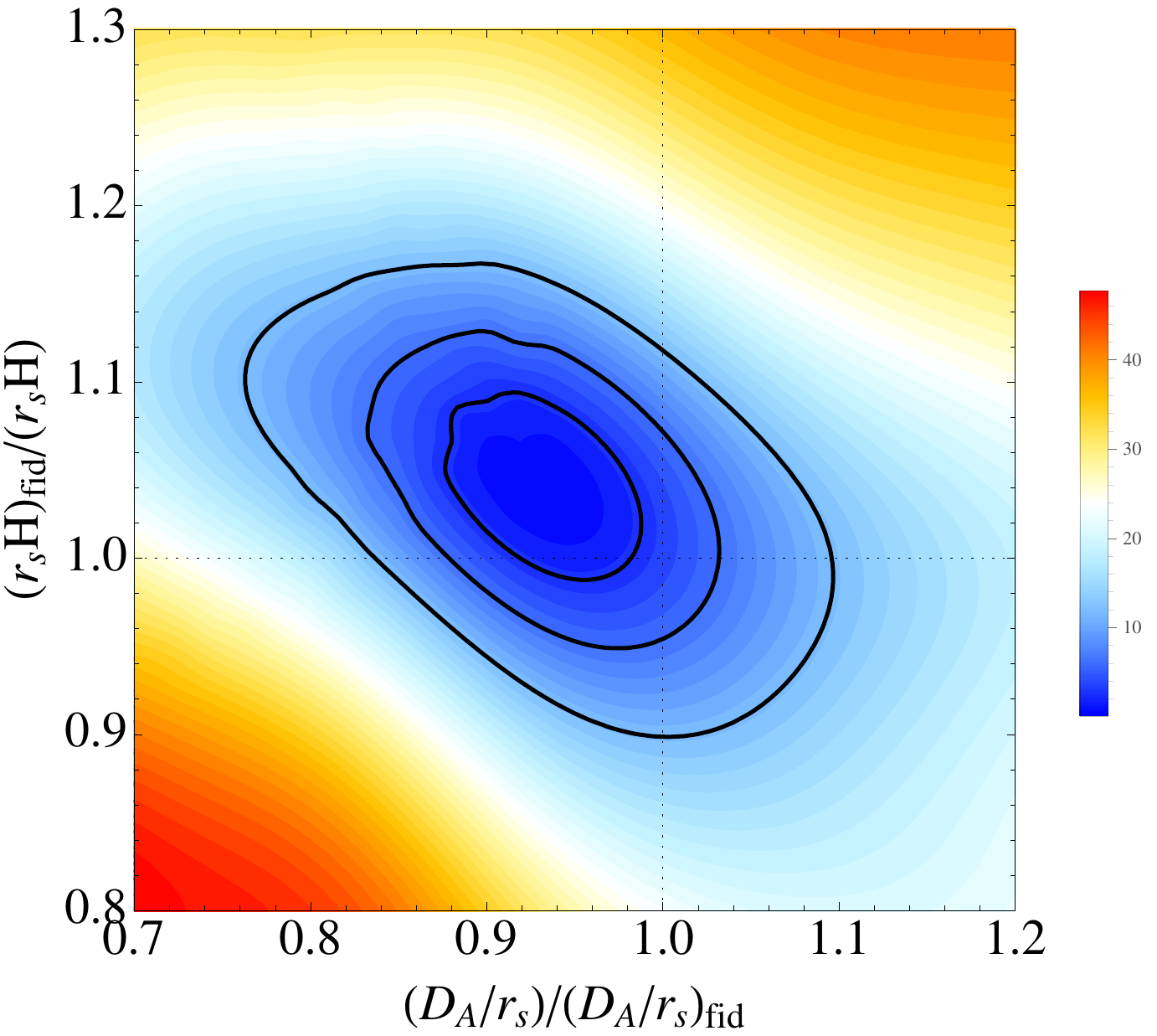}
 \end{center}
 \caption{$\Delta \chi^2$ as a function of $\apar$,$\aperp$
  (defined in equation \ref{eq:alpha}) in our fiducial analysis, 
  after marginalizing over the remaining 18 parameters. 
  The solid contours correspond to $\Delta \chi^2=2.27$, $5.99$ and $11.62$,
  equivalent to Gaussian probabilities of $68\%$, $95\%$ and $99.7\%$. 
  The fiducial model is consistent at the $\sim 1.5\sigma$ level.
}
 \label{fig:chi2}
\end{figure}

In figure \ref{fig:chi2} we present the main result of this paper: 
the value of $\Delta \chi^2$ as a function of ($\apar$,$\aperp$) for our 
fiducial BAO analysis, fully marginalized over the other 18 free parameters. 
The solid contours correspond to $\Delta \chi^2=2.27$, $5.99$ and $11.62$, 
equivalent to likelihood contours of $68\%$, $95\%$ and $99.7\%$ for a 
Gaussian likelihood. 
The fiducial model is consistent at the $\sim 1.5\sigma$ level.

We can translate our measurement of ($\apar$,$\aperp$) to a
measurement of the Hubble parameter and the angular diameter distance at
our mean redshift $z=2.36$, up to a factor $r_s$:
\begin{equation} 
 c/(H(z=2.36) ~ r_s) = 9.0 \pm 0.3 ~,  \qquad
 D_A(z=2.36)~ / ~ r_s = 10.8 \pm 0.4 ~.
\end{equation}

Using the best fit value of the sound horizon from the Planck collaboration
($r_s=147.49 \mpc$) \cite{2013arXiv1303.5076P}
\footnote{Table 2, column with 68\% limits for Planck+WP.}, 
we can present the results as:
\begin{equation}
 H(z=2.36) = 226 \pm 8 \kms \impc 	\qquad
 D_A(z=2.36) = 1590 \pm 60 \mpc ~.
\end{equation} 

\subsection{Systematic tests}
\label{ss:syst}

Table \ref{tab:bao_bb} presents the dependence of our results on the 
broadband model for our fiducial analysis.
We start by presenting the results in the absence of any broadband distortion
term (NO BB row), and we increasingly add more free parameters to our
model in an attempt to remove the distortion caused by the continuum
fitting method (see Figure 17 of \cite{2012JCAP...11..059F}).
Adding a single constant (BB\_0) does not improve the fit, but adding
a $r_\perp^{-1}$ term reduces the best fit $\chi^2$ by 30. The 
goodness of fit keeps improving while adding new free parameters, until
it saturates close to our fiducial model (BB\_7), after which adding
new parameters does not improve the fit much.
The BAO results are very insensitive to the chosen broadband function form
for all models with more than 6 free parameters (BB\_4 - BB\_14), and even 
the model with only 2 free parameters gives very similar results.

\begin{table}[h!]
 \centering
 \small
 \begin{tabular}{cc|cccc|c}
  Model & $N_{bb}$ & $\beta_F$  &  $b_q$  &  $\apar$  &  $\aperp$  & $\chi^2$ Prob \\
\hline
NO BB (-) & 0 			& $2.85 \pm 0.62 $  & $2.75 \pm 0.16 $  & $1.049 \pm 0.039 $  & $0.902 \pm 0.048 $  & 0.056 \\ 
BB\_0 (0, 0, 0, 0) & 1 		& $2.87 \pm 0.62 $  & $2.75 \pm 0.16 $  & $1.050 \pm 0.039 $  & $0.902 \pm 0.048 $  & 0.054 \\ 
BB\_1 (0, 0, -1, 0) & 2		& $1.55 \pm 0.30 $  & $2.73 \pm 0.16 $  & $1.038 \pm 0.039 $  & $0.921 \pm 0.041 $  & 0.255 \\
BB\_2 (0, 1, -1, 0) & 4		& $1.51 \pm 0.29 $  & $2.75 \pm 0.16 $  & $1.039 \pm 0.038 $  & $0.921 \pm 0.040 $  & 0.254 \\
BB\_3 (0, 2, -1, 0) & 6		& $1.22 \pm 0.25 $  & $2.94 \pm 0.18 $  & $1.047 \pm 0.036 $  & $0.916 \pm 0.040 $  & 0.318 \\
BB\_4 (0, 2, -2, 0) & 9 	& $0.91 \pm 0.22 $  & $3.11 \pm 0.20 $  & $1.044 \pm 0.035 $  & $0.927 \pm 0.037 $  & 0.350 \\
BB\_5 (0, 2, -2, 1) & 12	& $0.95 \pm 0.23 $  & $3.14 \pm 0.21 $  & $1.043 \pm 0.034 $  & $0.933 \pm 0.036 $  & 0.397 \\
BB\_6 (0, 2, -3, 0) & 12	& $0.94 \pm 0.24 $  & $3.05 \pm 0.22 $  & $1.047 \pm 0.035 $  & $0.924 \pm 0.037 $  & 0.381 \\ 
BB\_7 (0, 2, -3, 1) & 15  	& $1.09 \pm 0.29 $  & $3.02 \pm 0.22 $  & $1.042 \pm 0.034 $  & $0.930 \pm 0.036 $  & 0.404 \\ 
BB\_8 (0, 3, -2, 1) & 16	& $0.88 \pm 0.22 $  & $3.14 \pm 0.21 $  & $1.043 \pm 0.033 $  & $0.932 \pm 0.036 $  & 0.467 \\ 
BB\_9 (0, 3, -3, 1) & 20	& $1.01 \pm 0.27 $  & $3.01 \pm 0.22 $  & $1.042 \pm 0.034 $  & $0.929 \pm 0.036 $  & 0.463 \\
BB\_10 (-1, 3, -2, 1) & 20	& $0.89 \pm 0.22 $  & $3.14 \pm 0.21 $  & $1.045 \pm 0.033 $  & $0.931 \pm 0.036 $  & 0.443 \\ 
BB\_11 (0, 4, -2, 1) & 20	& $0.74 \pm 0.18 $  & $3.23 \pm 0.26 $  & $1.052 \pm 0.034 $  & $0.930 \pm 0.035 $  & 0.525 \\ 
BB\_12 (0, 5, -2, 1) & 24	& $0.80 \pm 0.23 $  & $3.17 \pm 0.27 $  & $1.051 \pm 0.034 $  & $0.932 \pm 0.035 $  & 0.517 \\ 
BB\_13 (0, 4, -2, 2) & 25	& $0.75 \pm 0.21 $  & $3.15 \pm 0.28 $  & $1.052 \pm 0.034 $  & $0.923 \pm 0.035 $  & 0.514 \\ 
BB\_14 (0, 4, -3, 1) & 25	& $0.90 \pm 0.27 $  & $3.03 \pm 0.28 $  & $1.047 \pm 0.035 $  & $0.929 \pm 0.036 $  & 0.513 \\ 
 \end{tabular}
 \caption{Best fit parameters for different broadband distortion models,
  defined by their values of ($i_{min}$,$i_{max}$,$j_{min}$,$j_{max}$) in 
  equation \ref{eq:bb}. $N_{bb}$ is the total of free parameters of the 
  broadband distortion model. The last column shows the probability inferred
  from the $\chi^2$ value in the fit and the number of degrees of freedom 
  (different in each row). 
  The first row (NO BB) shows the results for a fit without distortion, 
  and row BB\_7 corresponds to our fiducial model.}
 \label{tab:bao_bb}
\end{table}

Table \ref{tab:bao_range} presents the dependence on the separation range 
over which the cross-correlation is fitted, when the maximum separation is 
modified from the fiducial value of $180\hmpc$ to $170 \hmpc$ (RMAX\_170) 
or to $190 \hmpc$ (RMAX\_190), and the minimum separation from $40\hmpc$ 
to $30 \hmpc$ (RMIN\_30) or to $50 \hmpc$ (RMIN\_50).
The last three rows show the results of restricting the range of the angle
cosine $\mu = r_\parallel/r$ to $\| \mu \| < 0.8$ (MU\_08),
$\| \mu \| < 0.9$ (MU\_09), or $\| \mu \| < 0.95$ (MU\_095). The results in
this table show that the BAO measurement in general has little dependence 
on the fitting range. The broadband distortion is most important for
separations near the line of sight (i.e., $\| \mu \|$ near one), but the
removal of this most contaminated part does not significantly alter the
BAO peak position that is obtained, except in the MU\_095 case where the
position that is obtained shifts to a value closer to the expected one
in our fiducial model by nearly $1\sigma$.

\begin{table}[h!]
 \centering
  \begin{tabular}{c|ccccc}
   &  $\beta_F$  &   $b_q$  &  $\apar$  &  $\aperp$  & $\chi^2$ (d.o.f) \\
\hline
 FID         & $1.09 \pm 0.29 $  & $3.02 \pm 0.22 $  & $1.042 \pm 0.034 $  & $0.930 \pm 0.036 $  & 426.4 (420) \\ 
\hline
 RMAX\_170   & $1.07 \pm 0.28 $  & $3.07 \pm 0.22 $  & $1.046 \pm 0.034 $  & $0.926 \pm 0.036 $  & 397.4 (394) \\ 
 RMAX\_190   & $1.05 \pm 0.27 $  & $3.04 \pm 0.22 $  & $1.042 \pm 0.034 $  & $0.929 \pm 0.036 $  & 438.3 (436) \\ 
 RMIN\_30    & $1.03 \pm 0.21 $  & $2.85 \pm 0.14 $  & $1.043 \pm 0.035 $  & $0.930 \pm 0.038 $  & 443.5 (430) \\ 
 RMIN\_50    & $1.65 \pm 0.71 $  & $3.01 \pm 0.34 $  & $1.044 \pm 0.034 $  & $0.923 \pm 0.038 $  & 407 (406) \\ 
\hline
 MU\_08      & $2.26 \pm 2.0 $   & $2.61 \pm 0.38 $  & $1.025 \pm 0.088 $  & $0.937 \pm 0.055 $  & 257.1 (244) \\ 
 MU\_09      & $1.39 \pm 0.89 $  & $2.73 \pm 0.41 $  & $1.023 \pm 0.054 $  & $0.938 \pm 0.041 $  & 306.7 (302) \\ 
 MU\_095     & $0.88 \pm 0.32 $  & $3.13 \pm 0.27 $  & $1.009 \pm 0.041 $  & $0.949 \pm 0.037 $  & 353.6 (344) \\ 
 \end{tabular}
 \caption{Best fit parameters for the fiducial analysis (FID), 
  and different fitting ranges (in $\hmpc$). 
  The last rows show the results when using only bins that are far 
  from the line of sight, with $|\mu| < 0.8$ (MU\_08),
  $|\mu| < 0.9$ (MU\_09) and $|\mu| < 0.95$ (MU\_095).
}
 \label{tab:bao_range}
\end{table}

\subsection{Test of the covariance matrix}

In table \ref{tab:bao} we can see that the $\chi^2$ value in our fiducial fit
is good, in the sense that it is compatible with being drawn from a $\chi^2$ 
distribution with mean equal to the degrees of freedom in the problem, i.e., 
the number of bins used in the fit (440) minus the number of free 
parameters (20).

\begin{figure}
 \begin{center}
  \includegraphics[scale=0.6]{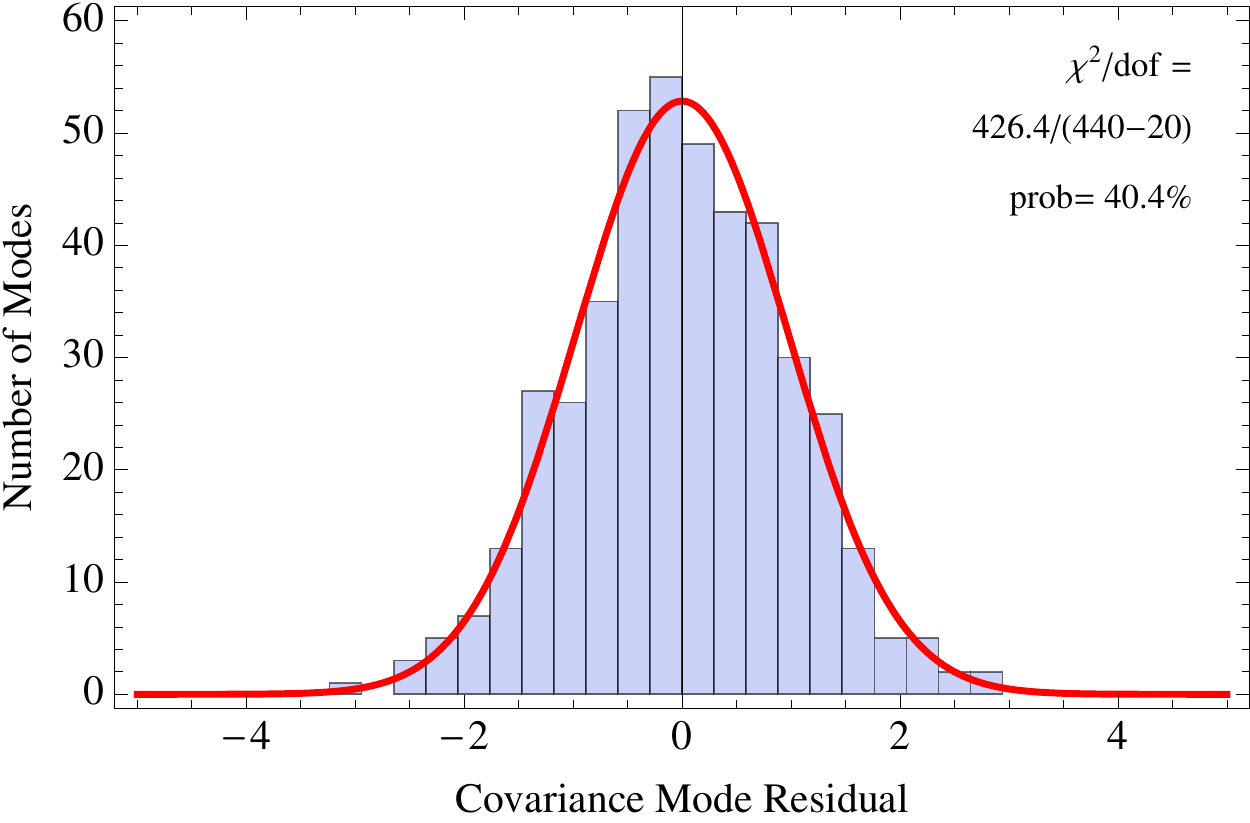}
 \end{center}
 \caption{Histogram of $\chi^2$ values for the different eigenmodes of the
  covariance matrix, compared to a zero-mean Gaussian with variance
  $(440-20)/440$, where 440 is the number of bins in the fit, and 20 is 
  the number of free parameters. The agreement between the distributions 
  supports the validity of our covariance matrix.
}
 \label{fig:cov}
\end{figure}

In order to test our estimate of the covariance matrix, we examine the 
distribution of $\chi^2$ for its different eigenmodes. The results of this
test are compared to a zero-mean Gaussian with variance $(440-20)/440$ in
figure \ref{fig:cov}. 
The agreement supports the validity of our covariance matrix.

\subsection{Alternative uncertainty estimates of the BAO scales}

The error on the fitted parameters reported so far have been computed
from the second derivatives of the log-likelihood function at its maximum,
assuming this likelihood function to be Gaussian at $1\sigma$. The BAO 
scale uncertainties in the fiducial analysis obtained in this way are 
$0.034$ for $\apar$ and 0.036 for $\aperp$. 
An alternative error estimate can be computed from the full likelihood 
surface in figure \ref{fig:chi2}, without assuming a Gaussian likelihood. 
The uncertainties obtained in the fiducial analysis are then $0.032$ 
for $\apar$ and $0.036$ for $\aperp$, in good agreement with the previous ones.

Both these estimates rely on the accuracy of the covariance matrix that we
have computed as described in section \ref{sec:cross}. We test this by 
computing an alternative bootstrap error on the BAO scale parameters, 
that does not rely on our covariance matrix.
We generate 1,000 bootstrap realizations of the survey \cite{Efron83}, 
combining the measurements from the 66 different sub-samples. 
The fitting analysis is done for each realization, and the uncertainties 
on $\apar$ and $\aperp$ are computed from their distribution of best fit values.
The resulting uncertainties on the BAO scales are $0.031$ on $\apar$ 
and $0.036$ on $\aperp$, in excellent agreement with the previous estimates. 

\subsection{Visualizing the BAO Peak}

Even though we do not use multipoles anywhere in our analysis, we present
here a fit of the multipoles from the measured cross-correlation in order to
better \textit{see} the BAO peak in the data.
We start by constructing a multipole expansion of our measured
cross-correlation, $\xi(r_\parallel,r_\perp)$, using a linear least-squares
fit to:
\begin{equation}
 \xi(r,\mu) = \sum_l L_l(\mu) \xi_l(r) ~,
\end{equation}
where $r = \sqrt{r_\parallel^2+r_\perp^2}$ and $\mu = r_\parallel / r$,
$L_l(x)$ is the Legendre polynomial of order $l$ and $\xi_l(r)$ are
the multipoles we wish to measure.

In figure \ref{fig:contours} we show the measured cross-correlation, as a
function of line of sight ($r_\parallel$) and transverse ($r_\perp$)
separation, together with our best fit model. From the right panel of the
figure, one can see that the best fit model of the broadband distortion is
asymmetric with respect to $r_\parallel=0$. Therefore we expect a net
non-zero contribution from odd multipoles.
For the purpose of visualization, however, we only fit the monopole
($l=0$) and the quadrupole ($l=2$), since these two multipoles contain most
of the cosmological infomation.
We use 36 equidistant interpolation points separated by $4 \hmpc$ and ranging
from $40 \hmpc$ to $180 \hmpc$.

Our estimates of the multipoles at different separations are highly
correlated. In order to improve the visualization of the BAO peak,
we apply a correction to the multipoles based on the analysis presented in
\cite{2013JCAP...03..024K}.
We start by examining the eigenmodes of the covariance matrix and identify
a particular mode being essentially a DC offset of the monopole, and therefore
responsible for much of the correlations between separations.
We then project out the mode from the data and its covariance matrix, and
refit for the distortion while keeping all other parameters fixed from the
baseline best fit.

\begin{figure}[h!]
 \begin{center}
   \includegraphics[scale=0.5]{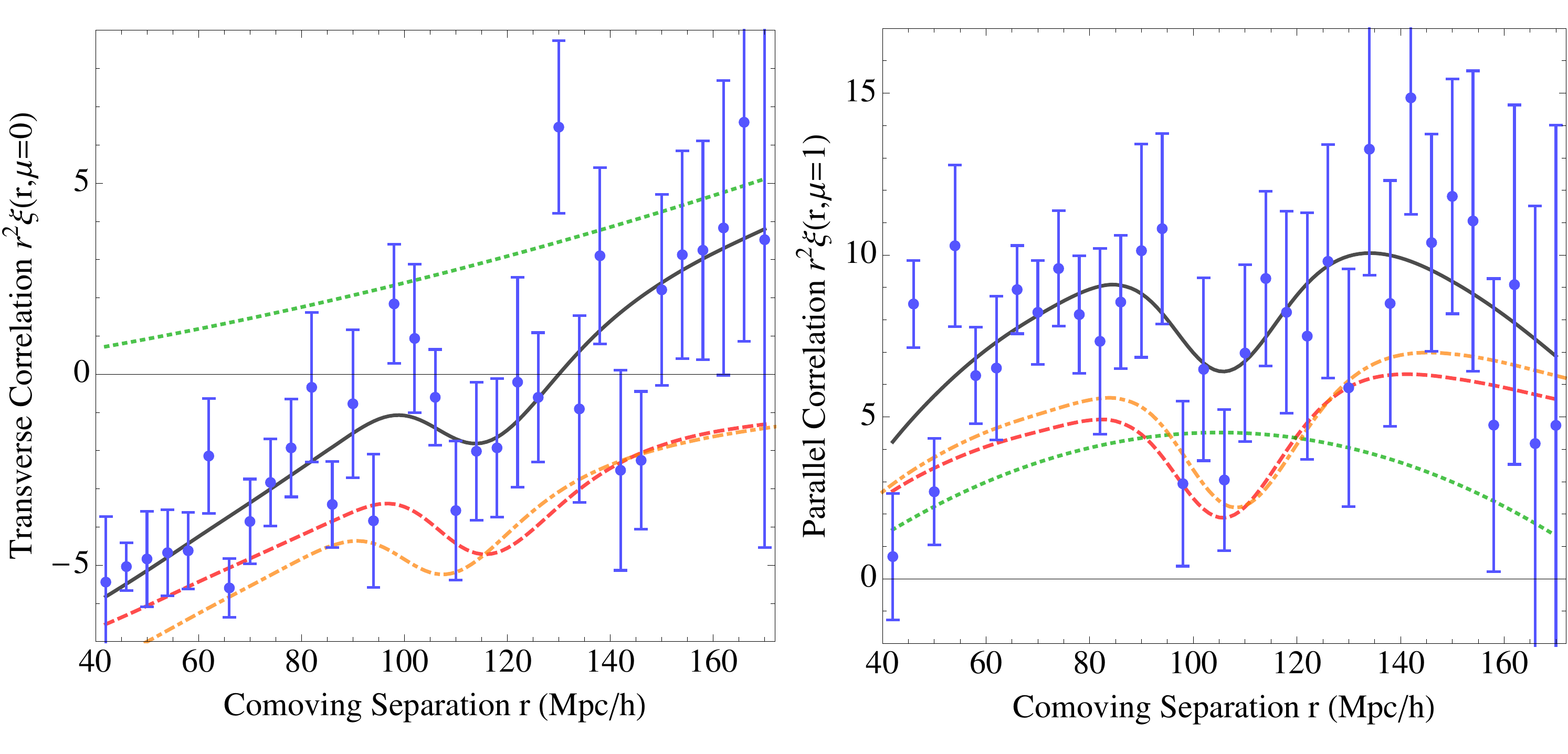}
 \end{center}
 \caption{Transverse (left) and parallel (right) correlations, defined as
  $\xi(r,\mu=0)=\xi_0(r)-\xi_2(r)/2$ and $\xi(r,\mu=1)=\xi_0(r)+\xi_2(r)$,
  after projecting out the mode responsible for most of the correlation
  between separations.
  The best fit theory is shown in a solid black curve, its BAO-only part in
  a red dashed curve and the distortion in a dotted green curve. The orange
  dot-dashed curve shows the cosmological signal for our fiducial cosmology
  ($\alpha=1$), using a quasar bias of $b_q=3.64$ and a \lya redshift-space
  distortion parameter $\beta_F=1.1$, as measured in
  \cite{2013JCAP...05..018F}.
  All datapoints and lines are weighted by $r^2$ and are plotted after the
  projection (see text for details).
 }
 \label{fig:mult}
\end{figure}

Figure \ref{fig:mult} shows the resulting monopole and quadrupole of the
quasar-\lya cross-correlation, expressed as the transverse correlation,
$\xi(r, \mu=0)=\xi_0(r)-\xi_2(r)/2$ (left panel), and the parallel
correlation $\xi(r, \mu=1)=\xi_0(r)+\xi_2(r)$.
We superimpose a fit with all parameters fixed from the
2D BAO fit except for the distortion. The solid black curve shows the best
fit, the red dashed curve is the BAO-only part (with parameters fixed from
the 2D fit), and the green dotted curve shows the distortion, which is
parabolic after $r^2$ weighting.
Since the \lya fluctuation is defined in equation \ref{eq:delta_f} as a
transmission fluctuation, positive values of $\delta_F$ reflect negative
density fluctuations, implying a negative value for the bias factor
$b_F$. This explains why the BAO feature in the quasar-\lya cross-correlation
appears as a dip instead of a peak, as seen in figure \ref{fig:mult}.

The orange curve in figure \ref{fig:mult} shows the predicted
cross-correlation for our fiducial cosmological model with
$\apar=\aperp=1$, and an amplitude determined by a quasar bias factor
$b_q=3.64$ and a \lya redshift distortion parameter $\beta_F=1.1$, as
measured in \cite{2013JCAP...05..018F}. The fact that this model is consistent
with the best fit that is obtained here to the DR11 data proves that our
result is consistent with that obtained in \cite{2013JCAP...05..018F} using
the DR9 data, and that the different values that are obtained in our fit
for $b_q$ and $\beta_F$ are caused by our addition of an arbitrary broadband
function, with parameters that are degenerate with $b_q$ and $\beta_F$.
The amplitude of the BAO dip, as visualized in figure \ref{fig:mult}, is
consistent with our expectation. This is seen also in the model AMP in
table \ref{tab:bao}, where the parameter $a_{\rm peak}$ has a best fit value
that is consistent with unity. A model with a suppressed BAO peak (model NW
in table \ref{tab:bao}) has a $\chi^2$ that is worse than our fiducial model 
by $20$, although we warn that this is not to be directly interpreted as a
statistical significance of a BAO detection because our likelihood function
is not necessarily Gaussian. In any case, our interest here lies in the
statistical constraint obtained on the BAO scale, rather than the
significance of the BAO detection in the quasar-\lya cross-correlation only.

\section{Discussion \& Conclusions}
\label{sec:conc}

We have presented a measurement of the quasar - \lya cross-correlation using   
approximately 164,000 quasars from the eleventh Data Release (DR11) of SDSS. 
We are able to measure the BAO scale along and across the line of 
sight ($\alpha_\parallel$, $\alpha_\perp$) with an uncertainty 
of $3.4\%$ and $3.6\%$ respectively. The measurement is in agreement with 
our fiducial cosmology well within the $95\%$ confidence level. 

We have checked the robustness of our measurement under changes of 
broadband models, separation range used, and different error estimates. 
As discussed in section \ref{ss:MTC}, we are not particularly careful in 
our treatment of the non-BAO part of the cross-correlation. However, 
the best fit values for the bias parameters of both quasars and \lyaf\ 
are roughly consistent with previous analyses, with rather large 
uncertainties since we only use large separations to measure the BAO scale.

In table \ref{tab:comp} we compare the results with other BAO measurements
at the same redshift from the \lya auto-correlation measured with DR9 
(\cite{2013A&A...552A..96B}, \cite{2013JCAP...04..026S}).
We also present our results when using only data from DR10. 
Assuming that the uncertainties in these BAO measurements scale with the 
inverse of the square root of the survey area, we can extrapolate them from
DR9 to DR11, using $\sqrt{A_{DR11}/A_{DR9}}=1.60$.
We show these extrapolations in the last two rows of table \ref{tab:comp}. 

\begin{table}[h!]
 \centering
  \begin{tabular}{ccc|ccc}
  Analysis & Probe & Data Release & $\alpha$ & $\alpha_\parallel$ & $\alpha_\perp$ \\
 \hline
  Busca\_2013     & Auto  & DR9  & $1.01 \pm 0.03$  & -		       & - 	          \\
  Slosar\_2013    & Auto  & DR9  & $0.98 \pm 0.020$ & $0.99 \pm 0.035$ & $0.98 \pm 0.070$ \\
  This work       & Cross & DR10 & $1.00 \pm 0.027$ & $1.06 \pm 0.038$ & $0.91 \pm 0.041$ \\ 
  This work       & Cross & DR11 & $0.99 \pm 0.022$ & $1.04 \pm 0.034$ & $0.93 \pm 0.036$ \\ 
 \hline
  Busca\_2013     & Auto  & to DR11 & $\pm 0.019$ &    -        &     -       \\
  Slosar\_2013    & Auto  & to DR11 & $\pm 0.013$ & $\pm 0.022$ & $\pm 0.046$ \\
 \end{tabular} 
 \caption{Comparison of different BAO analysis at $z \sim 2.4$ from BOSS, 
  from the auto-correlation in DR9 (Busca\_2013 \cite{2013A&A...552A..96B}, 
  Slosar\_2013 \cite{2013JCAP...04..026S}), and from this work. 
  We show our results when using only DR10 data and when including DR11 data. 
  In the last two rows we extrapolate the uncertainties of previous work to DR11, 
  assuming that these scale with the inverse of the square root of the survey area.}
 \label{tab:comp}
\end{table}

The errors on the BAO scale ($\apar$,$\aperp$) from our DR10 analysis are 
considerably smaller than those reported in \cite{OConnell2013}. 
An extensive comparison of the two analyses within the BOSS \lya working 
group concluded that the discrepancy can be explained by the differences in 
the analysis. While \cite{OConnell2013} uses only the monopole and the 
quadrupole to fit the BAO scale, in this analysis we use the full 2D
contours of the cross-correlation function.

In the absence of any broadband distortion of $\xi(r_\parallel,r_\perp)$ 
(or with a distortion that is a-priori known), we find that essentially all of 
the BAO signal is contained within the monopole $\xi_0(r)$ and quadrupole 
$\xi_2(r)$. 
However, when broadband distortion is present, as in our analysis, it 
contributes significantly to multipoles other than the monopole and quadrupole,
and leads to correlated uncertainties between distortion and BAO parameters and 
corresponding parameter degeneracies.  
As a result, we find that the unknown broadband distortion parameters can be 
determined more precisely with a fit to the full $\xi(r_\parallel,r_\perp)$ 
(or, equivalently, a larger set of multipoles) instead of a fit to only the 
monopole and quadrupole. 
Similarly, we find that a fit to $\xi(r_\parallel,r_\perp)$ yields a more 
precise determination of the BAO parameters by helping to break the degeneracy 
between distortion and BAO parameters. 
The actual improvement we find is a factor of 1.2 in $\apar$ and a factor of 1.3
in $\aperp$. \cite{DR11BAO} found that a similar improvement is also seen 
when measuring BAO from the \lya auto-correlation function, although a detailed
study on mock data sets revealed a large scatter in the gain from 
realization to realization.

\subsection{\lya auto-correlation vs. quasar-\lya cross-correlation}

In appendix \ref{app:forecast} we present a Fisher matrix projection comparing
the relative strength of measuring BAO with the \lya auto-correlation and with
the quasar-\lya cross-correlation. In a BOSS-like survey, both probes should
measure the transverse BAO scale with similar uncertainties, while the \lya
auto-correlation should be able to measure the line of sight scale
$\sim 40\%$ better than the cross-correlation with quasars.

The measurement of BAO from the \lya auto-correlation in DR9 was presented 
in \cite{2013A&A...552A..96B} and \cite{2013JCAP...04..026S}. Most of the 
difference between the uncertainties in these results can be explained by 
the looser data cuts used in \cite{2013JCAP...04..026S}, that included lines 
with DLAs and that defined their \lya forest with a wider wavelength range. 
In this analysis we used data cuts similar to those in 
\cite{2013JCAP...04..026S}, and therefore we compare here our uncertainties 
with those from \cite{2013JCAP...04..026S} extrapolated to DR11 
(see table \ref{tab:comp}). 

Our measurement of $\apar$ is $\sim 55\%$ worse than the results from the \lya 
auto-correlation of \cite{2013JCAP...04..026S} extrapolated to DR11, in 
good agreement with the prediction of $\sim 40 \%$ computed in the appendix. 
The Fisher forecast formalism predicted similar uncertainties in $\aperp$, 
and we find that our measurement is $\sim 20 \%$ better than the extrapolated 
results from the auto-correlation.

In the same appendix we also show that on the scales of interest for
BAO measurements ($k > 0.05 \ihmpc$) cosmic variance is not the dominant
contribution to our error budget. Assuming that the shot noise in the quasar
density field is uncorrelated with the small scale fluctuations in the \lya 
absorption and with the instrumental noise, we can then combine both BAO
measurements as if they were independent.

\begin{figure}
 \begin{center}
  \includegraphics[scale=0.3]{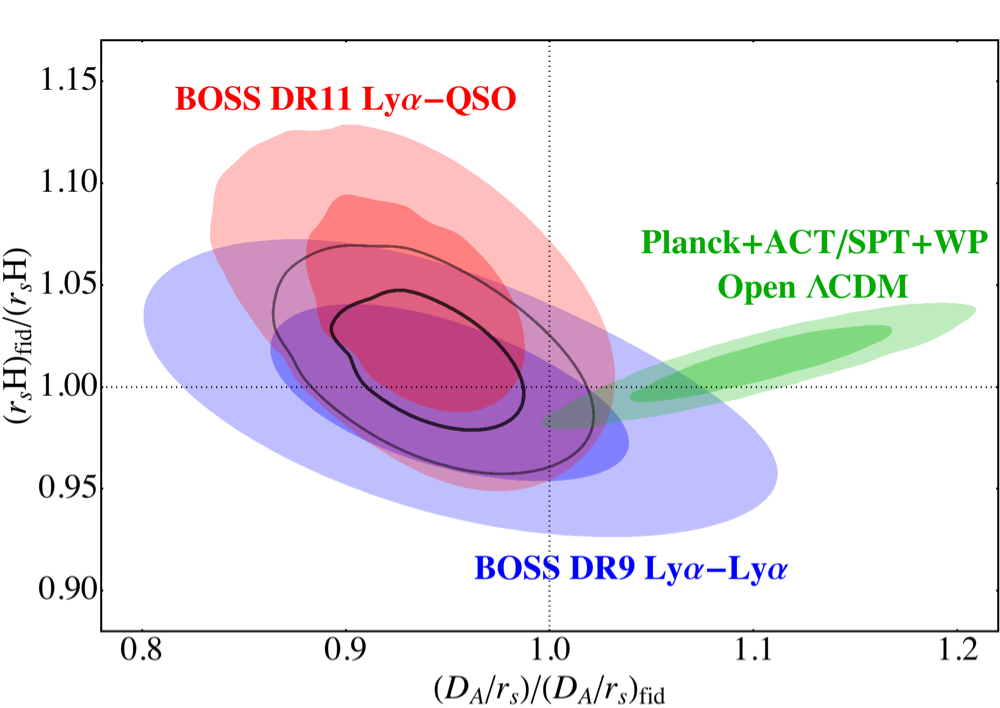}
 \end{center}
 \caption{Contours of $\Delta \chi^2=2.27$ and $5.99$, corresponding to 
  Gaussian confidence levels of 68\% and 95\%, 
  from the \lya auto-correlation analysis from DR9
  (\cite{2013JCAP...04..026S}, in blue), from the cross-correlation
  from DR11 (this work, in red) and from the joint analysis
  (in black). The green contours show the 68\% and 95\% contours 
  for the regions of this parameter space allowed by the Planck results 
  \cite{2013arXiv1303.5076P} in an open $\Lambda$CDM cosmology.
 }
 \label{fig:joint}
\end{figure}

In figure \ref{fig:joint} we compare the contours on ($\apar$,$\aperp$) from
the \lya auto-correlation function from DR9 (\cite{2013JCAP...04..026S}
in blue, generated from the files in 
\url{http://darkmatter.ps.uci.edu/baofit/}), and compare it to our measurement 
from the cross-correlation function from DR11 (red) and the sum of their 
$\chi^2$ surfaces (in black), assuming they are independent. 
We compare these constraints with the 68\% and 95\% confidence 
limits obtained from the Planck results \cite{2013arXiv1303.5076P} in 
an open $\Lambda$CDM cosmology, shown in green.
\footnote{We use the Planck + ACT/SPT + WP public chains available under 
the name base\_omegak\_planck\_lowl\_lowLike\_highL.}
Note that by allowing for space curvature, the Planck constraints 
on the distance and expansion rate at our mean redshift $z=2.36$ are much 
less restrictive compared to a flat model.

We have shown that adding the cross-correlation of \lya and quasars to the 
auto-correlation of \lya can certainly improve the constraints on BAO scales 
at high redshfit. 
A detailed analysis of the cosmological implications of the measurements of 
the \lya auto-correlation and the quasar-\lya cross-correlation will be 
presented in a future publication, which will include the DR11 results from 
the \lya auto-correlation, together with a more complete examination of 
potential correlations between the two measurements.

\begin{acknowledgments}

We would like to thank Ross O'Connell for detailed comparisons with his 
analysis, and Shirley Ho for very useful comments. 

This research used resources of the National Energy Research Scientific 
Computing Center (NERSC), which is supported by the Office of Science of 
the U.S. Department of Energy under Contract No. DE-AC02-05CH11231.
DK would like to thank CEA Saclay for their hospitality and productive 
environment during his sabbatical.
JM is supported in part by Spanish grant AYA2012-33938.

Funding for SDSS-III has been provided by the Alfred P. Sloan
Foundation, the Participating Institutions, the National Science
Foundation, and the U.S. Department of Energy Office of Science.
The SDSS-III web site is http://www.sdss3.org/.

SDSS-III is managed by the Astrophysical Research Consortium for the
Participating Institutions of the SDSS-III Collaboration including the
University of Arizona,
the Brazilian Participation Group,
Brookhaven National Laboratory,
University of Cambridge,
Carnegie Mellon University,
University of Florida,
the French Participation Group,
the German Participation Group,
Harvard University,
the Instituto de Astrofisica de Canarias,
the Michigan State/Notre Dame/JINA Participation Group,
Johns Hopkins University,
Lawrence Berkeley National Laboratory,
Max Planck Institute for Astrophysics,
Max Planck Institute for Extraterrestrial Physics,
New Mexico State University,
New York University,
Ohio State University,
Pennsylvania State University,
University of Portsmouth,
Princeton University,
the Spanish Participation Group,
University of Tokyo,
University of Utah,
Vanderbilt University,
University of Virginia,
University of Washington,
and Yale University.

\end{acknowledgments}

\bibliography{cosmo,cosmo_preprints}
\bibliographystyle{revtex}

\appendix
\section{Public Access to Data and Code}
\label{app:access}

The baofit software used in this paper is publicly available at 
\url{https://github.com/deepzot/baofit/}.
The measured cross-correlation function and its covariance matrix, and 
the instructions to reproduce the BAO constraints presented 
in this paper, can be downloaded from 
\url{http://darkmatter.ps.uci.edu/baofit/}, together with the likehood 
surface used to generate Figure \ref{fig:chi2}.
The software is written in C++ and uses MINUIT \cite{1975CoPhC..10..343J} for 
likelihood minimization. 

\section{Fisher Matrix Forecasts}
\label{app:forecast}

The goal of this appendix is to compare the expected accuracy with which one 
should be able to measure the BAO scale from a BOSS-like survey using the
\lya auto-correlation function and the quasar-\lya cross-correlation. 

Instead of undertaking a full forecast of the uncertainty on the BAO scale, 
we will do a simpler comparison and evaluate the signal to noise ratio (S/N) 
that one should obtain for a certain Fourier mode $\vk$. 
This is a fair comparison, since the uncertainty on the BAO scale should be 
proportional to the uncertainty on the power spectrum over the relevant BAO 
scales.

\subsection{Auto-correlation}

We start by computing the expected signal to noise ratio in the 
auto-correlation of \lya and in the auto-correlation of quasars. On large 
scales, the signal in the auto-correlation can be described with a simple 
linear bias model with the Kaiser model to account for redshift space 
distortions:
\begin{equation}
 P_A(\vk) = b_A^2 \left(1 + \beta_A \mu^2 \right)^2  ~ P(k) ~,
\end{equation}
where $b_A$ and $\beta_A$ are the linear bias parameter of the tracer $A$ and 
its redshift space distortion parameter, $P(k)$ is the matter power spectrum,
and $\mu$ is the cosine of the angle between the Fourier mode $\vk$ and the 
line of sight. 

The accuracy with which one can measure the quasar power spectrum 
$P_A(\vk)$ in a given bin centered at ($k$,$\mu$) 
can be quantified by the signal to noise ratio (S/N),
\begin{equation}
 \left( \frac{S}{N} \right)^2_A 
     = N_k ~\frac{P^2_A(\vk)} {var[P_A(\vk)]} ~,
\end{equation} 
where $N_k$ is the number of modes in the bin. Since we only care about
relative performance in this appendix, we will drop any $N_k$ and will plot
signal to noise ratio per mode. 

For a sample of point-like sources (for instance quasars), the variance of
its measured power spectrum can be approximated by
\begin{equation} 
 var \left[ P_A(\vk) \right] = 2 \left(P_A (\vk) 
                + n_A^{-1}\right)^2 ~,
\end{equation}
with $n_A$ the number density of systems. 

Since the \lya forest is not a discrete point sampling of the underlying 
matter density field, but rather a non-linear transformation of a continuous 
sampling along discrete lines of sight, we need to use a slightly different 
approach. \cite{2007PhRvD..76f3009M} 
computed the expected S/N in the measurement of $P_{FF}(\vk)$ in a 
spectroscopic survey, and highlighted the importance of the 
`` aliasing term '' due to the sparse sampling of the universe. 
Here we use the formalism from \cite{2011MNRAS.415.2257M} that combines both 
the noise term and the aliasing term defining a noise-weighted density 
of lines of sight per unit area $n_{eff}$, 
\begin{equation}
  var[P_{FF}(\vk)] = 2 \left(P_{FF}(\vk) + P^{1D}(k \mu) ~
                                   n_{eff}^{-1} \right)^2 ~ ,
\end{equation} 
where $P^{1D}(k \mu)$ is the one-dimensional flux power spectrum.

\subsection{Cross-correlation}

The cross correlation between the \lya absorption and the quasar density field
can be defined as 
\begin{equation}
 \left< \delta_F(\vk) ~ \delta_q(\vk \prime) \right> 
  = (2 \pi)^3 \delta^D(\vk+\vk \prime) ~ P_{qF}(\vk) ~.
\end{equation}
Again, in the linear regime we can relate the cross-correlation power 
spectrum with the linear power spectrum $P(k)$ using the linear bias 
parameters defined above, 
\begin{equation}
 P_{qF}(\vk) = b_q \left(1+\beta_q \mu^2 \right) 
           ~ b_F \left(1+\beta_F \mu^2 \right) P(k) ~.
 \label{eq:cross}
\end{equation}
\cite{2011MNRAS.415.2257M} showed that the variance in the measurement 
of the cross-correlation can be approximated by
\begin{equation}
  var \left( P_{qF}(\vk) \right) = 
      P_{qF}(\vk)^2 + \left(P_{qq}(\vk) + n_q^{-1} \right) 
      \left(P_{FF}(\vk) + P^{1D}(k \mu) ~ n_{eff}^{-1} \right) ~.
\end{equation}
In this approximation, the expected S/N in a bin of ($k$,$\mu_k$) can be
approximated by
\begin{equation}
 \left( \frac{S}{N} \right)^2_{Fg} 
     = N_k ~ \frac{P^2_{qF}(\vk)} 
            {P_{qF}(\vk)^2 + \left(P_{qq}(\vk) + n_q^{-1}\right) 
                \left(P_{FF}(\vk) + 
                   P^{1D}(k \mu) ~ n_{eff}^{-1} \right) } ~ .
\end{equation}

\subsection{Forecast for a BOSS-like survey}

Here we quantify the previous results for the case of a spectroscopic survey
with properties similar to the BOSS survey. The BOSS survey has an area 
of $A=10^4 \, {\rm deg}^2$, and if we restrict the analysis to the redshift 
range $2 < z < 3$, the total volume of the survey is roughly 
$V=40 (\hgpc)^3$.
The quasar density in the BOSS survey is roughly 
$n_q = 160 000 / V \sim 4 \times 10^{-6} (\hmpc)^{-3}$, 
and we assume a quasar bias of $b_q=3.6$ 
(\cite{2013JCAP...05..018F},\cite{2012MNRAS.424..933W}).
The effective density of lines of sight for BOSS is estimated in 
\cite{2011MNRAS.415.2257M}
to be $n_{eff} \approx 10^{-3} (\hmpc)^{-2}$, and we assume the 
values for the \lya biases of $b_F = -0.15$ and $\beta_F = 1.2$, 
both compatible with the 1D measurement of \cite{MCSEET06} and the 
3D clustering from \cite{2011JCAP...09..001S}.
We compute the power spectra at our fiducial redshfit of $z_c = 2.36$.

\begin{figure}[h!]
  \begin{center}
    \begin{tabular}{cc}
      \includegraphics[scale=0.5, angle=-90]{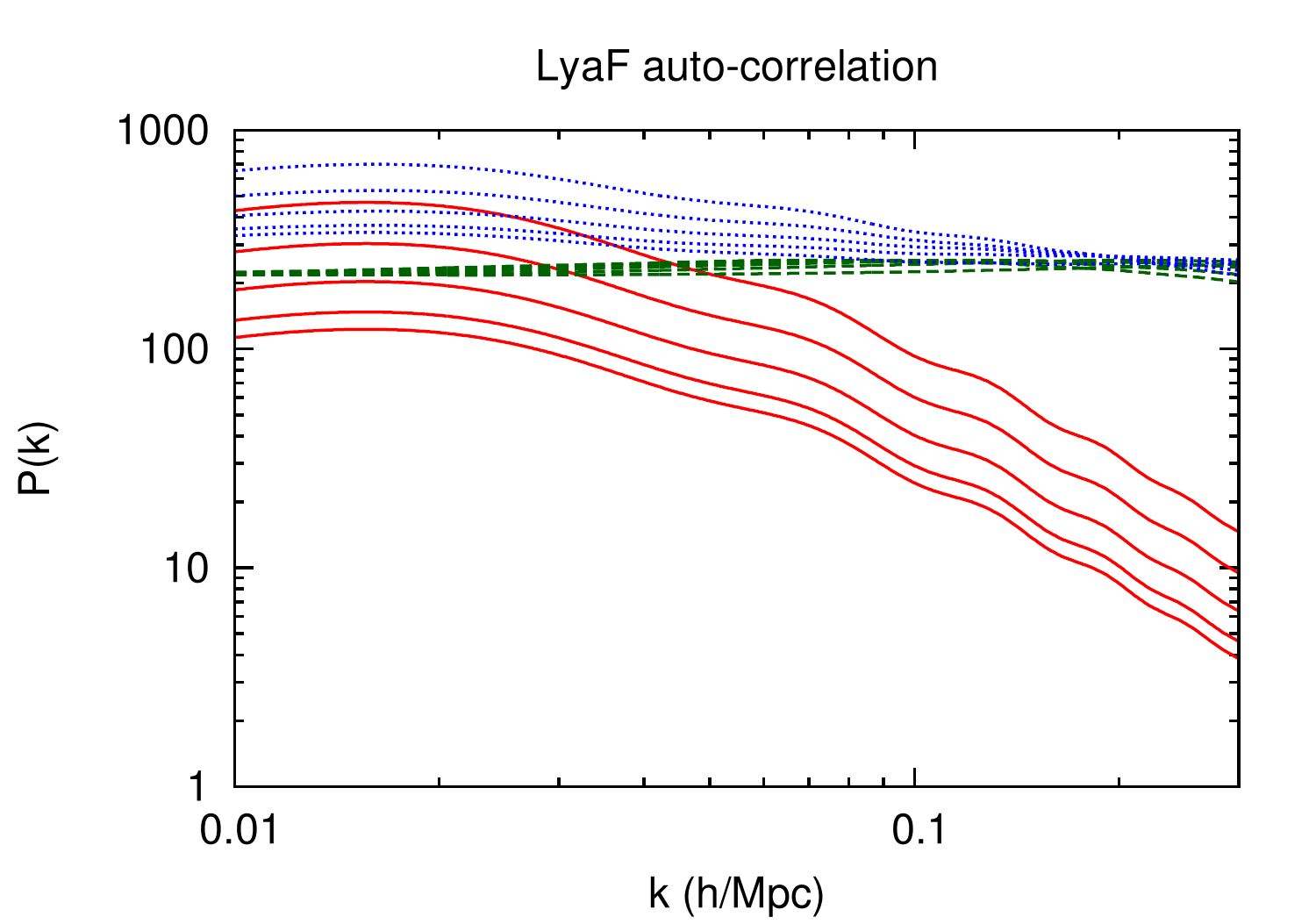} &
      \includegraphics[scale=0.5, angle=-90]{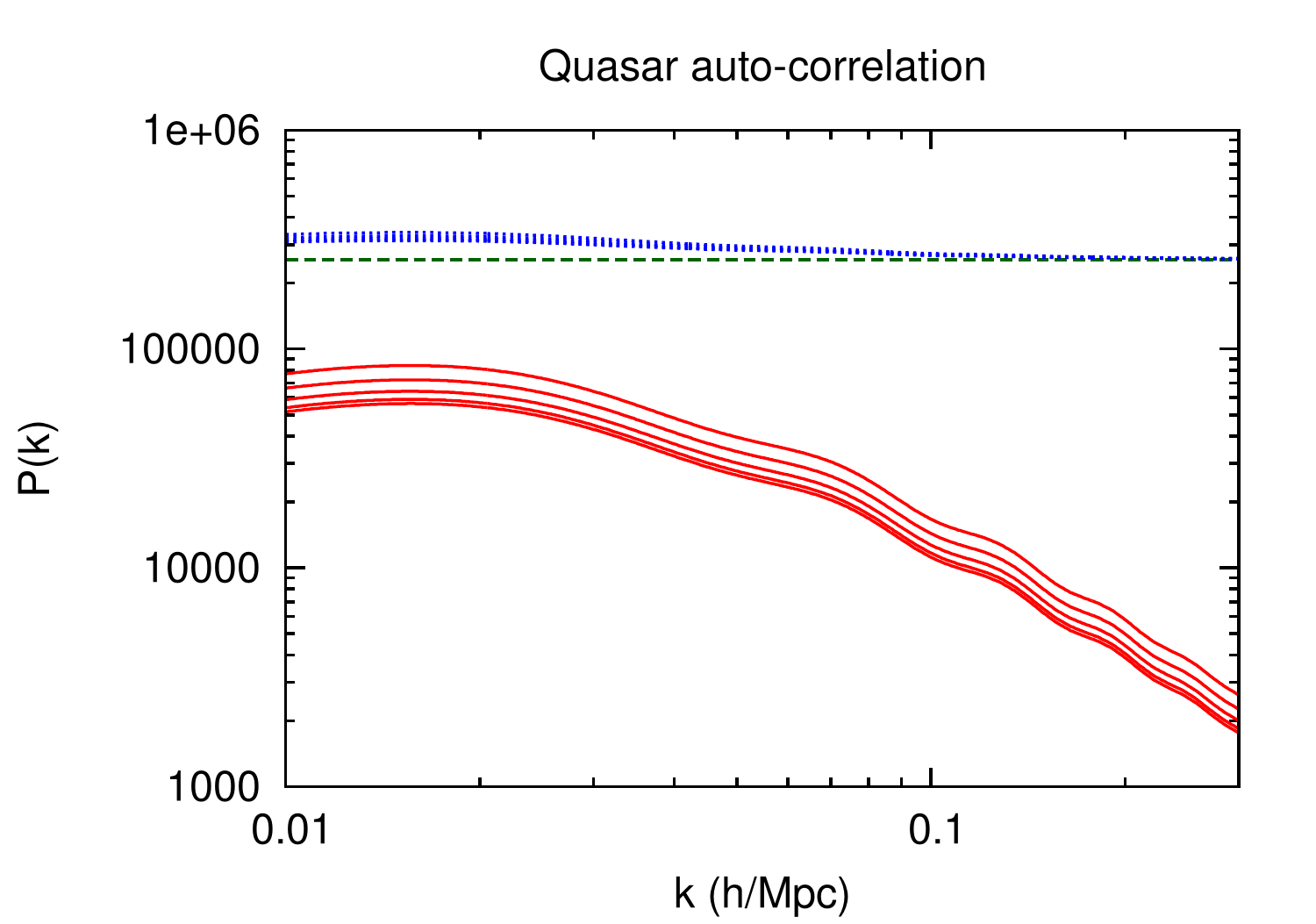} \\
      \includegraphics[scale=0.5, angle=-90]{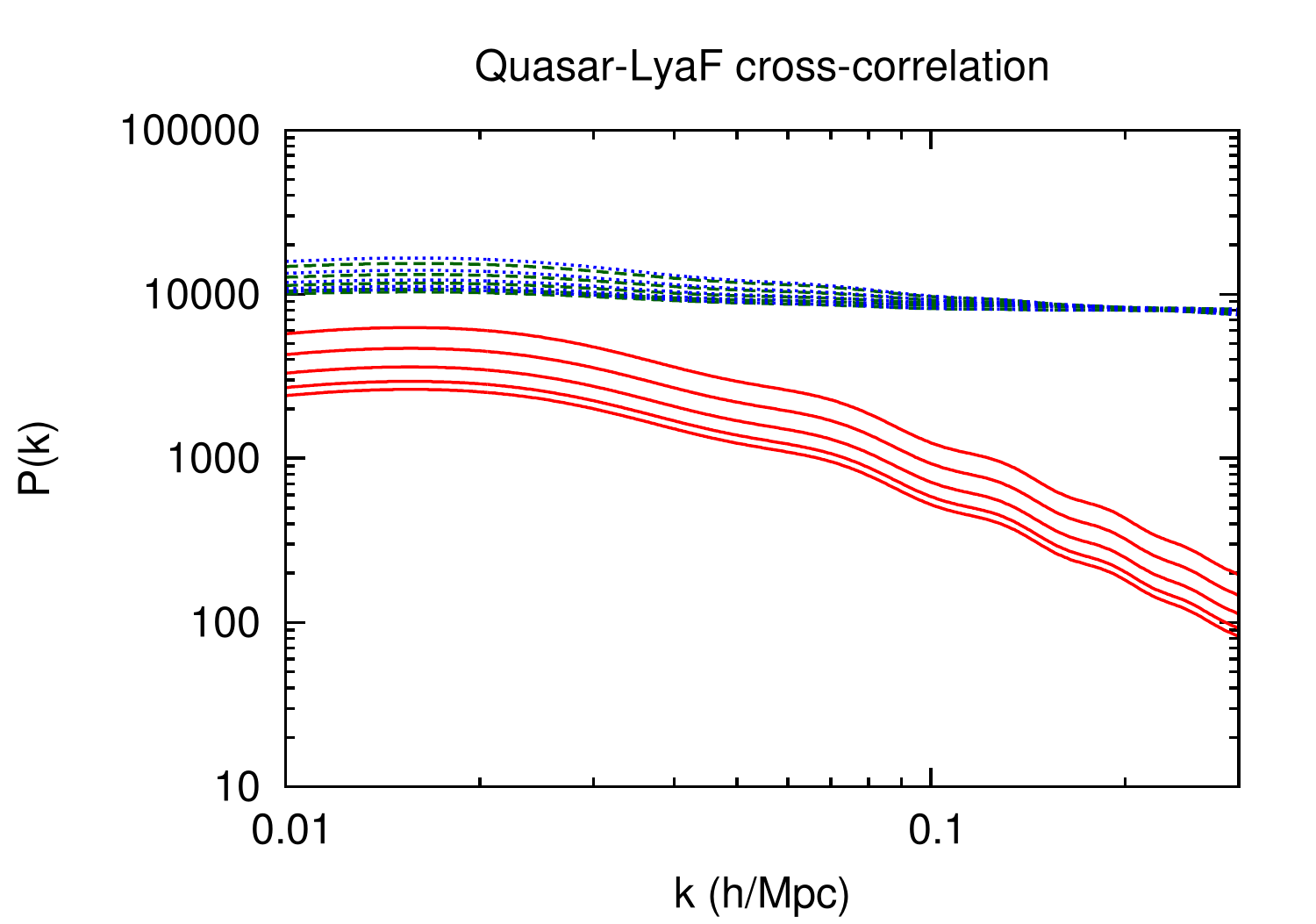} & 
      \includegraphics[scale=0.5, angle=-90]{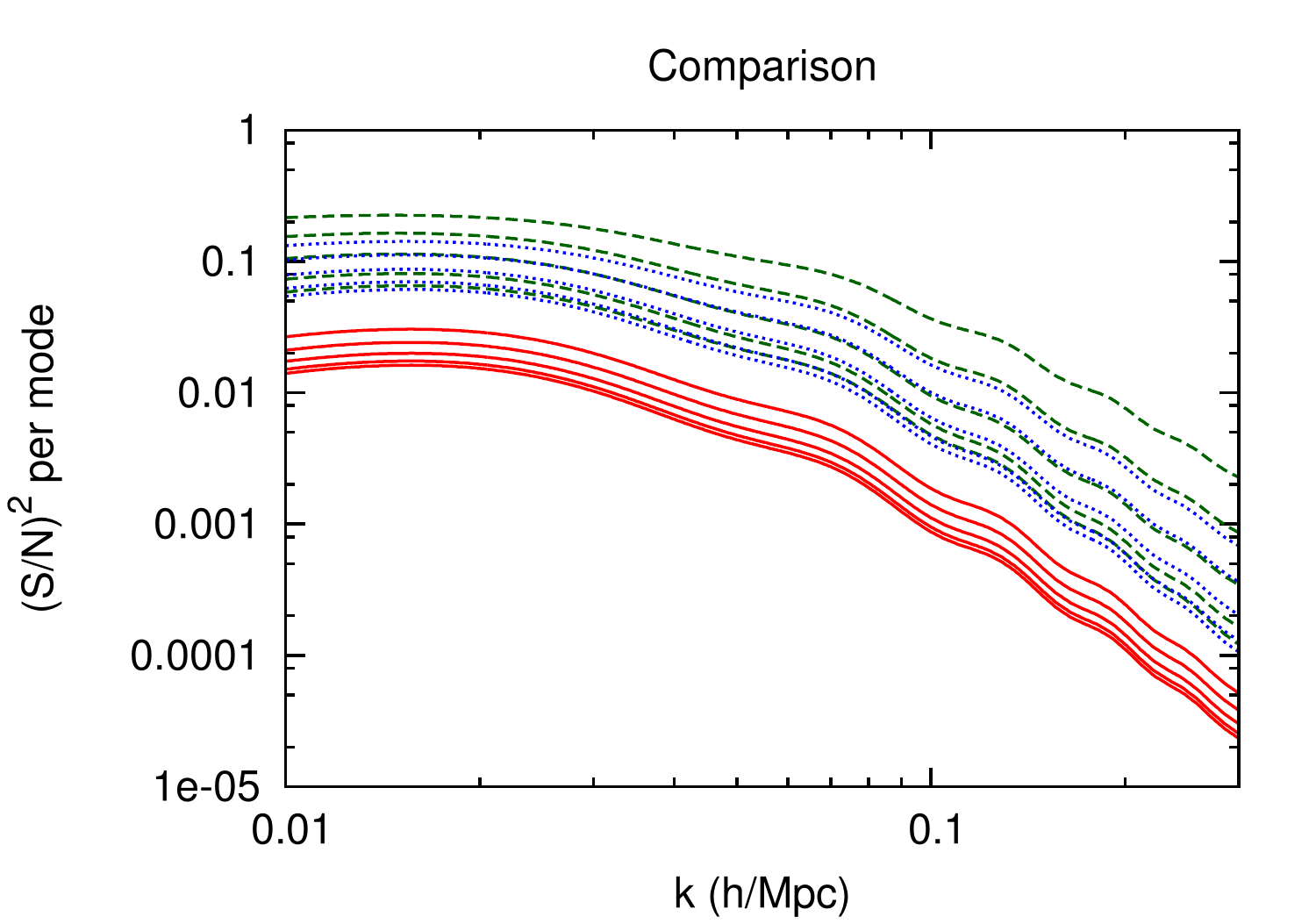} \\
   \end{tabular}
 \end{center}
 \caption{Study of the signal to noise ratio in different analyses:
  auto-correlation of \lya (top left), auto-correlation of quasars 
  (top right) and their cross-correlation (bottom left). 
  The red solid lines show the signal, the green dashed lines the 
  ``shot noise'' level and the dotted blue lines their sum, for different 
  values of $\mu$ (increasing from lower to upper lines).
  The bottom-right panel shows the expected $(S/N)^2$ per mode for the 
  \lya auto-correlation (dashed green), quasar auto-correlation (solid red) 
  and cross-correlation (dotted blue). 
 }
 \label{fig:fisher}
\end{figure}

In figure \ref{fig:fisher} we compare the signal and the different noise
contributions for the different analyses: \lya auto-correlation (top left), 
quasar auto-correlation (top right) and quasar-\lya cross-correlation 
(bottom left). In the bottom-right panel we compare the expected signal to 
noise ratio (squared) per mode for the three different analyses, and for 
different values of $\mu_k$.
We can see that the S/N of the quasar-\lya cross-correlation is much higher 
than the quasar auto-correlation, and that for transverse modes (lower lines)
is as high as the \lya auto-correlation.
It is also clear from the figure that on scales relevant for BAO 
($k>0.05\ihmpc$), we are in the noise-dominated regime, and therefore 
cosmic variance is at best a secondary contribution to the the error budget.

\end{document}